\documentclass[review,authoryear]{elsarticle}

\usepackage[T1]{fontenc}
\usepackage{lmodern}   
\usepackage{textcomp}  

\usepackage[colorlinks=true,allcolors=blue]{hyperref}
\usepackage{amsmath,amssymb,amsfonts}
\usepackage{graphicx}
\usepackage{xcolor}
\usepackage[all]{nowidow}
\usepackage[inline]{enumitem}
\usepackage[scaled]{beramono}
\usepackage{booktabs} 
\usepackage[colorinlistoftodos]{todonotes}
\usepackage[tight,footnotesize]{subfigure}
\usepackage{multirow}
\usepackage{balance}
\usepackage{listings}
\usepackage{cleveref}
\usepackage{soul}
\usepackage{pdflscape}
\usepackage{lineno}
\usepackage{color, colortbl}
\usepackage{csquotes}
\usepackage[width=12cm, left=4.8cm]{geometry}
\usepackage{amssymb}
\usepackage{xspace}
\usepackage{fontawesome}
\usepackage{tikz,pifont}
\usepackage{tikz}
\usetikzlibrary{positioning,arrows.meta}
\usepackage[many]{tcolorbox}
\usepackage{xcolor, soul}
\usepackage{xcolor}
\usepackage{listings}

\definecolor{lightyellow}{rgb}{1,1,0.7}
\sethlcolor{lightyellow}

\lstdefinestyle{pythonstyle}{
  language=Python,
  basicstyle=\ttfamily\scriptsize,
  keywordstyle=\color{blue}\bfseries,
  stringstyle=\color{red!60!brown},
  commentstyle=\color{green!50!black}\itshape,
  showstringspaces=false,
  breaklines=true,
  frame=single,
  rulecolor=\color{black!30},
  numbers=none,
  captionpos=t
}

\colorlet{lightgray}{gray!30}
\sethlcolor{lightgray}

\DeclareUnicodeCharacter{00A0}{~}
\def\BibTeX{{\rm B\kern-.05em{\sc i\kern-.025em b}\kern-.08em
		T\kern-.1667em\lower.7ex\hbox{E}\kern-.125emX}}

\definecolor{mygreen}{rgb}{0,0.6,0}
\definecolor{mygray}{rgb}{0.95,0.95,0.95}
\definecolor{myred}{rgb}{0.5,0,0}

\lstdefinestyle{JavaStyle} {
	backgroundcolor=\color{white},   
	commentstyle=\color{mygreen}, 
	breakatwhitespace=false,
	keywordstyle=\color{violet},
	language=Java,
	stringstyle=\color{blue},
	basicstyle=\scriptsize\ttfamily,
	showstringspaces=false }

\newcommand{\mybox}[4]{
	\begin{figure}[h]
		\centering
		\begin{tikzpicture}
			\node[anchor=text,text width=\columnwidth-0.5cm, draw, rounded corners, line width=1pt, fill=#3, inner sep=2mm] (big) {\\#4};
			\node[draw, rounded corners, line width=.3pt, fill=#2, anchor=west, xshift=3mm] (small) at (big.north west) {#1};
		\end{tikzpicture}
	\end{figure}
}

\newcommand*{\tool}{\texttt{LAMPS}\@\xspace}

\usepackage{amssymb}

\newcommand*{\ie}{i.e.,\@\xspace}

\newcommand*{\MP}{\texttt{MPHunter}\@\xspace}

\makeatletter
\newcommand*{\etc}{%
	\@ifnextchar{.}%
	{etc}%
	{etc.\@\xspace}%
}
\makeatother
\newcommand*{\etal}{et~al.\@\xspace}
\newcommand\revised[1]{\textcolor{black}{#1}}

\definecolor{verylightgray}{gray}{0.99}
\definecolor{lightgray}{gray}{0.92}

\definecolor{mygreen}{rgb}{0,0.6,0}
\definecolor{mygray}{rgb}{0.95,0.95,0.95}
\definecolor{myred}{rgb}{0.5,0,0}

\lstdefinestyle{JavaStyle} {
	backgroundcolor=\color{verylightgray},   
	commentstyle=\color{mygreen}, 
	breakatwhitespace=false,
	keywordstyle=\color{violet},
	language=Java,
	stringstyle=\color{blue},
	basicstyle=\scriptsize,
	showstringspaces=false
}

\lstdefinestyle{searchstringstyle}{
	basicstyle=\ttfamily\scriptsize,
	captionpos=t,                    
	numbers=none,                    
	numbersep=5pt,                  
	showspaces=false,                
	showstringspaces=false,
	showtabs=false,                  
	tabsize=2,
	frame=single
}

\newcommand{\shadedtext}[1]{\texttt{\hl{\small #1}}}
\definecolor{darkgreen}{rgb}{0.0, 0.5, 0.0}

\newtcolorbox{shadedbox}{
	drop shadow southeast,
	breakable,
	enhanced jigsaw,
	colback=white,
	boxrule=0.80pt,
	left=0.3em,
	right=0.3em,
	top=0.1em,
	bottom=0.05em
}

\newcommand*{\FA}{\texttt{Fetcher Agent}\@\xspace}
\newcommand*{\EA}{\texttt{Extractor Agent}\@\xspace}
\newcommand*{\CA}{\texttt{Classifier Agent}\@\xspace}
\newcommand*{\VA}{\texttt{Verdict Agent}\@\xspace}

\newcommand{\rqone}{\revised{\textbf{RQ$_1$}: \emph{How does \tool compare with unsupervised \MP and supervised TF-IDF-based stacking model?}}}

\newcommand{\rqtwo}{\textbf{RQ$_2$}: \emph{How does \tool compare with RAG-based configurations?}} 

\newcommand{\rqthree}{\textbf{RQ$_3$}: \emph{Is the deployment of multi LLMs-based agents advantageous with respect to only one agent?}}

\begin{document}
	\begin{frontmatter}

	\title{Many Hands Make Light Work: An LLM-based Multi-Agent System for Detecting Malicious PyPI Packages}

	\author[univaq]{Muhammad Umar Zeshan}
	\author[univaq]{Motunrayo Ibiyo}
	\author[univaq]{Claudio Di Sipio}
	\author[univaq]{\\Phuong T. Nguyen}
	\author[univaq]{Davide Di Ruscio\corref{cor1}}	
	\address[univaq]{Universit\`a degli studi dell'Aquila, 67100 L'Aquila, Italy\\
		\{\href{mailto:muhammadumar.zeshan@student.univaq.it}{muhammadumar.zeshan}, \href{mailto:motunryoosatohanmen.ibiyo@student.univaq.it}{motunryoosatohanmen.ibiyo}\}@student.univaq.it, \{\href{mailto:claudio.disipio}{claudio.disipio}, \href{mailto:phuong.nguyen@univaq.it}{phuong.nguyen}, \href{mailto:davide.diruscio@univaq.it}{davide.diruscio}\}@univaq.it}
	\cortext[cor1]{Corresponding author}
	


\begin{abstract}
Malicious code in open-source repositories such as PyPI poses a growing threat to software supply chains. Traditional rule-based tools often overlook the semantic patterns in source code that are crucial for identifying adversarial components. Large language models (LLMs) show promise for software analysis, yet their use in interpretable and modular security pipelines remains limited. 

This paper presents \tool, a multi-agent system that employs collaborative LLMs to detect malicious PyPI packages. The system consists of four role-specific agents for \textit{package retrieval}, \textit{file extraction}, \textit{classification}, and \textit{verdict aggregation}, coordinated through the CrewAI framework. A prototype combines a fine-tuned CodeBERT model for classification with LLaMA~3 agents for contextual reasoning. \tool has been evaluated on two complementary datasets: D$_1$, a balanced collection of 6,000 \texttt{setup.py} files, and D$_2$, a realistic multi-file dataset with 1,296 files and natural class imbalance. On D$_1$, \tool achieves 97.7\% accuracy, \revised{surpassing \MP and TD-IDF stacking models--two state-of-the-art approaches.} On D$_2$, it reaches 99.5\% accuracy and 99.5\% balanced accuracy, outperforming RAG-based approaches and fine-tuned single-agent baselines. McNemar's test confirmed these improvements as highly significant. The results demonstrate the feasibility of distributed LLM reasoning for malicious code detection and highlight the benefits of modular multi-agent designs in software supply chain security.

	\end{abstract}
	\end{frontmatter}


	%

\section{Introduction}


Open-source platforms foster software development by enabling programmers to share and access extensive collections of reusable code \citep{DBLP:books/sp/rsse2014}. This practice helps developers accelerate their work and increase productivity. Unfortunately, \emph{every rose has its thorn}: the availability of a considerable amount of reusable components also creates opportunities for adversarial users to inject malicious code into widely used libraries, posing significant security risks to software supply chains \citep{10.1145/3691620.3695492,10.1145/3510003.3510146}. Malicious packages are deliberately disguised as legitimate code to trick users, making detection particularly challenging \citep{alfadel2023empirical}. As a result, recognizing malicious code is becoming increasingly critical, especially with the proliferation of AI-based systems that are trained on publicly available repositories.

A variety of approaches have been developed to automatically analyze security threats, including static rule-based scanners~\citep{9647791}, signature-driven classifiers~\citep {Chakraborty2020262}, clustering methods \citep{10298315}, and behavior sequence models~\citep{10.1145/3705304}. While these methods achieve encouraging performance, they often fail in the presence of obfuscation, indirect API usage, or logic-level concealment. The recent success of large language models (LLMs) has opened new opportunities for analyzing and reasoning about software systems. Their ability to interpret code semantics through natural language prompts makes them attractive for tasks such as code summarization, defect detection, and vulnerability identification \citep{chen2021evaluating,ahmad2021unified}. Despite this potential, applications in security-critical domains such as malicious code detection in open-source repositories remain limited. Moreover, many existing attempts treat LLMs as monolithic black boxes, where a single model instance is tasked with multiple stages of analysis without meaningful modularity or transparency \citep{wang2023can,ohm2020backstabber}. Empirical evidence on Retrieval Augmented Generation (RAG) further indicates that retrieval alone does not reliably improve the detection of malicious intent in PyPI packages \citep{ibiyo2025detectingmalicioussourcecode}. Altogether, this motivates the development of an approach that can reliably surface file-level signals and provide auditable package-level decisions.

Recently, the software engineering community has witnessed the emergence of multi-agent systems that leverage orchestrations of LLMs to decompose complex tasks into specialized agents, often surpassing single-agent pipelines \citep{10.1145/3712003,nguyen2025teamworkmakesdreamwork}. While these systems show promise for coding tasks, there remains a need to investigate coordination patterns, the use of shared context, and the feasibility of LLM-based multi-agent designs for security analysis.

In this work, we propose \tool, a novel \textbf{\underline{L}}LM-based multi \textbf{\underline{A}}gent system for detecting \textbf{\underline{M}}alicious \textbf{\underline{P}}yPI Package\textbf{\underline{S}}. We decompose the detection task into a coordinated network of specialized agents with clearly defined roles and prompts, orchestrated through CrewAI \citep{taulli2025crewai}. Instead of a monolithic pipeline, responsibilities such as package retrieval, file filtering, code-level classification, and verdict aggregation are distributed across autonomous units that communicate through structured prompt-based messages. \CA uses a fine-tuned CodeBERT model trained on labeled Python files, while the other agents rely on LLaMA-3 for role-specific reasoning. This design reflects a shift from singular model invocation to structured, distributed reasoning, and it is supported by a clarified threat and detection scope that focuses on learned code-level indicators.

Although this work explores a new direction, it is based on our previous work on the detection of malicious packages~\citep{ibiyo2025detectingmalicioussourcecode}. We further extend the original framework using an LLM-based multi-agent system, and conducting an empirical evaluation with real-world datasets. 
We describe agent roles and communications, construct and validate datasets for two settings, adopt a rigorous evaluation protocol with package-level splits and repeated runs, test statistical significance with McNemar’s test, and report standard metrics alongside balanced accuracy in imbalanced settings. We also record efficiency measurements on our local environment to complement accuracy-based comparisons. \tool has been evaluated against recent baselines, including \MP~\citep{10298315} and RAG-based configurations~\citep{ibiyo2025detectingmalicioussourcecode}, using the same datasets and experimental settings. 

The key contributions of this paper are as follows:

\begin{itemize}
	
	\item \textbf{Solution.} We introduced \tool, a multi-agent LLM-based system for detecting malicious code in open-source Python repositories. The design leverages role specialization for package retrieval, file filtering, and file-level classification with a fine-tuned CodeBERT, as well as conservative package-level aggregation.
	
	\item \textbf{Evaluation.} Using two datasets with 6,000 and \revised{1,296 samples}, \revised{we evaluated \tool against state-of-the-art baselines, including \MP~\citep{10298315}, a TF-IDF-based stacking model~\citep{samaana2025machine} working on top of machine learning and static analysis, 
	and RAG-based approaches~\citep{ibiyo2025detectingmalicioussourcecode}}, following an explicit protocol with package-level splits, repeated runs with fixed seeds, and significance testing. Through the evaluation, we also demonstrate that the application of LLM-based multi-agent systems is advantageous compared to that of single LLMs.
	
	\item \textbf{Open Science.} We released a replication package with full source code, data preparation scripts, prompts, and environment specifications to facilitate independent verification \citep{lampsjss2025}.
	
\end{itemize}

\noindent
\noindent
\textbf{Structure.} The remainder of this paper is organized as follows. Section~\ref{sec:background} presents the background and motivation for the study. Section~\ref{sec:methodology} describes the design of \tool, including agent responsibilities, coordination, and threat model. Section~\ref{sec:poc} provides the proof of concept, which also includes dataset construction. Section~\ref{sec:ExperimentalResult} presents the experimental setup, results for three research questions, and a discussion of findings. Section~\ref{sec:related} reviews related work. Finally, Section~\ref{sec:conclusion} concludes the paper and outlines directions for future research.

\section{Motivation and Background}
\label{sec:background}

This section introduces the problem context and highlights the motivations that guided the design of our approach. We first describe how malicious code typically manifests in Python package ecosystems and why it poses unique detection challenges. Afterwards, two motivating examples are introduced to illustrate how hidden patterns could hinder the detection of malicious PyPI packages. We then outline the role of LLMs 
in capturing semantic properties of code, and discuss how a multi-agent architecture provides practical and conceptual benefits over a monolithic pipeline. Together, these elements establish the foundation for the methodology detailed in the next section.

\subsection{Motivating examples}

The Python Package Index (PyPI) is a widely used repository that allows developers to distribute source archives containing both library code and installation scripts. A distinctive characteristic of this ecosystem is the reliance on files such as \texttt{setup.py}, which can execute arbitrary Python code during installation. This flexibility, while powerful for developers, creates an attack surface that adversaries can exploit by embedding harmful logic inside otherwise legitimate-looking packages.

Standard attacker techniques include typosquatting on popular package names to trick users into installing a malicious variant, inserting encoded strings to conceal commands, invoking \texttt{subprocess} calls to fetch and execute remote payloads, and performing file or network operations during installation or at first import. Because such behaviors are interleaved with benign functionality, they are difficult to detect through simple pattern matching or metadata checks.

A motivating example is shown in Listing~\ref{lst:malicious-example}, taken from the package \texttt{pong\-replace-10.4}, where the \texttt{setup.py} script constructs a base64-encoded command and executes it via a hidden \texttt{subprocess} call. The obfuscation ensures that surface-level similarity measures cannot easily distinguish the script from benign installers. This illustrates why detection must move beyond syntactic features to capture semantic cues of malicious behavior.

\begin{lstlisting}[style=pythonstyle,
	caption={Excerpt from the malicious \texttt{setup.py} of package \texttt{pongreplace-10.4}, showing the obfuscated payload execution.},
	label={lst:malicious-example}]
	import subprocess, os
	if not os.path.exists('tahg'):
	subprocess.Popen(
	'powershell -WindowStyle Hidden -EncodedCommand '
	'cABvAHcAZQByAHMAaABlAGwAbAAgAEkAbgB2AG8AawBlAC0AVwBlA...',
	shell=False,
	creationflags=subprocess.CREATE_NO_WINDOW)
\end{lstlisting}

\begin{figure}[h!]
	\centering
	\includegraphics[width=0.88\linewidth]{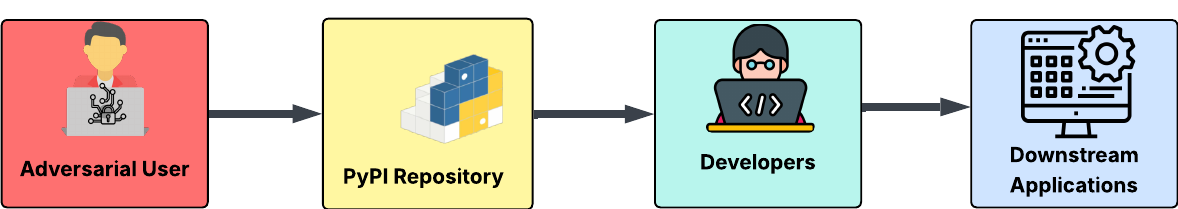}
	\caption{Motivating example of supply-chain risk in PyPI. An adversarial upload can embed obfuscated logic in \texttt{setup.py} or auxiliary modules; developers unknowingly install the package, and downstream applications are affected at runtime. The example motivates semantics-aware analysis and conservative package-level aggregation.}
	\label{fig:motivating-example}
\end{figure}

Figure~\ref{fig:motivating-example} complements this example by visualizing the supply-chain path from an adversarial upload to downstream compromise and highlighting the points where malicious logic can hide in installation scripts or auxiliary modules. The figure motivates the need for semantics-aware, auditable analysis that preserves file-level signals and supports package-level decisions.

Another observation is that malicious logic is not always confined to installation scripts. In multi-file packages, payloads can be distributed across auxiliary modules, with only a small fraction of files exposing suspicious patterns. A detector that operates only at the file level risks underestimating package-level risk.

\newpage
\begin{lstlisting}[style=pythonstyle,
	caption={Obfuscated malicious code in \texttt{esqpeppywvirtual-9.5}.},
	label={lst:malicious-example2}]
	from distutils.core import setup
	
	try:
	import subprocess
	import os
	if not os.path.exists('tahg'):
	# obfuscated payload (base64 encoded command)
	subprocess.Popen(
	'powershell -WindowStyle Hidden -EncodedCommand <truncated>',
	shell=False,
	creationflags=subprocess.CREATE_NO_WINDOW
	)
	except:
	pass
	
	try:
	setup(
	name='pongreplace',
	packages=['modlib'],
	version='10.4',
	description='A library for creating a terminal user interface',
	author='EsqueleSquad',
	author_email='tahgoficial@proton.me',
	classifiers=[
	'Programming Language :: Python :: 3'
	],
	)
	except:
	pass
\end{lstlisting}

Listing~\ref{lst:malicious-example2} shows a malicious \texttt{setup.py} script extracted from the PyPI package named \texttt{esqpeppywvirtual-9.5}. The script conditionally invokes a hidden PowerShell command\footnote{The PowerShell payload in Listing~\ref{lst:malicious-example2} has been truncated for the sake of presentation. The full scripts are available in the online appendix~\citep{lampsjss2025}.} that downloads and executes a remote payload, thereby compromising the host system during installation. This payload is obfuscated using a \shadedtext{base64-encoded} command and executed silently to evade detection. The package is disguised as a typical utility library, making it difficult to detect using traditional rule-based tools. 

\MP~\citep{10298315} is a state-of-the-art approach in detecting malicious PyPI packages, and we used it to test with the code in Listing~\ref{lst:malicious-example2}. The result showed that \MP fails to detect the hidden malicious intent, \ie it cannot recognize the silently \shadedtext{base64-encoded} command. This is consistent with a clustering-based mechanism that groups samples by surface-level similarity: obfuscated, single-file installer logic may not exhibit the frequency-based characteristics required to form a clear outlier cluster. 

These examples motivate us to come up with an approach that can recognize the hidden intent by means of a series of LLM-based agents. This leads to the conservative aggregation strategy in our proposal: 
if any file is flagged as malicious, the package is treated as malicious. Such a policy reduces false negatives at the ecosystem level, where the cost of a missed detection is typically higher than the cost of a false alarm.

These considerations shape the design of our pipeline: \emph{(i)} provenance-preserving retrieval and deterministic extraction to ensure that the correct inputs are analyzed; \emph{(ii)} file-level classification using a semantics-aware model capable of recognizing encoded payloads, process invocation, and unauthorized I/O; and \emph{(iii)} package-level aggregation to translate heterogeneous file-level outputs into an auditable decision. These elements are elaborated in Section \ref{sec:methodology} and empirically evaluated in Section \ref{sec:poc}.

\subsection{Large Language Models for Code Understanding}

LLMs have recently demonstrated strong capabilities in capturing the semantic properties of both natural and programming languages. When trained on large-scale code corpora, such models can embed source code into vector spaces that preserve syntactic structure and semantic intent, making them suitable for downstream tasks such as code summarization, defect detection, and vulnerability identification \citep{allamanis2018survey,chen2021evaluating}. Unlike traditional feature engineering or rule-based approaches, LLMs learn distributed representations that generalize across projects and programming idioms.

For our purposes, the following two observations are critical:

\begin{enumerate}
	\item \textit{General-purpose LLMs alone are not sufficient to distinguish malicious from benign code reliably.} Prior work shows that zero-shot or few-shot prompting without domain-specific adaptation often leads to unstable or superficial predictions in security-sensitive contexts. This limitation motivates the use of a model pre-trained on source code and then fine-tuned on a curated dataset of benign and malicious files. In \tool, we adopt CodeBERT as the classifier backbone because it has been explicitly pre-trained for program understanding and can be adapted through supervised training to capture behavioral signatures such as encoded payload decoding, process invocation, and unauthorized I/O.
	
	\item \textit{While file-level classifiers are necessary for detecting fine-grained malicious patterns, they must operate within broader workflows where multiple files and package metadata interact.} Models like LLaMA-3 offer complementary strengths, as they are capable of prompt-based reasoning and flexible input handling, making them well-suited for orchestrating auxiliary tasks such as package retrieval, file filtering, and verdict explanation. By combining a fine-tuned model specialized for classification (CodeBERT) with role-specific reasoning agents powered by a general-purpose LLM (LLaMA-3), our architecture balances accuracy at the classification step with flexibility in pipeline coordination.
	
\end{enumerate}

This design choice addresses a central challenge in malicious package detection: the need to preserve semantic sensitivity at the code level while enabling modular and auditable reasoning across the entire package. The resulting system leverages LLMs not as a monolithic solution but as specialized components integrated into a coordinated workflow, a strategy that we evaluate in detail in Section \ref{sec:ExperimentalResult}.

\subsection{Multi-Agent Orchestration and Motivation}

Although LLMs offer the semantic sensitivity needed for code analysis, relying on a single model instance as an end-to-end detector introduces both practical and conceptual limitations. A monolithic prompt that ingests all package content must simultaneously retrieve archives, extract files, classify source code, and provide justifications, without guarantees that each step is executed reliably. In multi-file packages, this approach is further constrained by context window limits, token truncation, and order sensitivity: malicious logic can be overlooked if a critical file is omitted or its signals diluted among benign content. For example, a package may contain dozens of benign utility modules alongside a \texttt{setup.py} script that silently downloads and executes a payload. A monolithic model truncated to fit within the context window may exclude the \texttt{setup.py} file entirely, misclassifying the package as benign. In contrast, a multi-agent system ensures that such critical files are always extracted, classified, and considered in the final verdict.

A multi-agent architecture mitigates these challenges by decomposing the overall workflow into specialized roles. Separate agents for fetching, extraction, classification, and verdict aggregation operate with clearly defined responsibilities and interact through structured prompts. This separation brings several advantages. First, it reduces the burden on individual models: the classifier can focus exclusively on semantic indicators of malicious code, while upstream agents ensure that inputs are curated and contextually valid. Second, it improves reproducibility and auditability, since intermediate outputs such as extracted file lists, per-file rationales, and aggregated verdicts can be logged and inspected, which is not possible in a monolithic black box pipeline. Third, it supports conservative package-level aggregation: when malicious logic is confined to a single file, the system still issues a cautious but transparent verdict.

The orchestration also enables extensibility. New roles, such as agents for dependency risk scoring or dynamic analysis, can be integrated without modifying the classifier or retraining the entire model. By framing malicious package detection as a distributed reasoning problem rather than a single-step classification, the system gains robustness against obfuscation and flexibility for adaptation. These motivations lead directly to the design of \tool, introduced in the Section \ref{sec:methodology}, which instantiates a coordinated network of LLaMA-3 and CodeBERT agents to realize this architecture.
	
\section{Methodology}
\label{sec:methodology}

\begin{figure}[!t]
	\centering
	\includegraphics[width=0.9\linewidth]{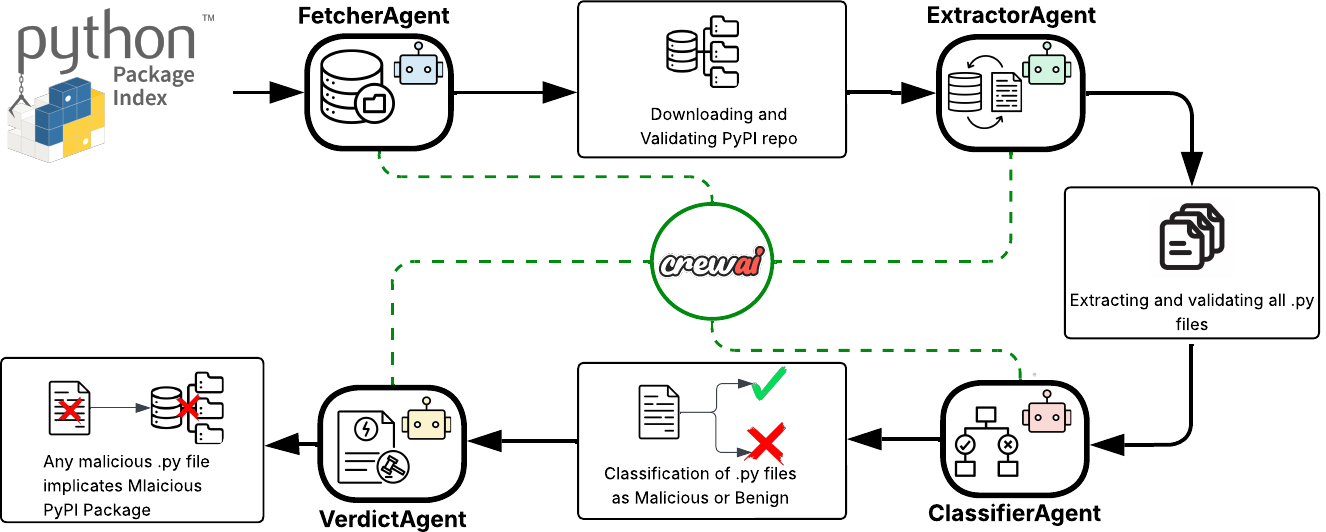}
	\caption{Architecture of \tool{} showing agent responsibilities and communication flow.}
	\label{fig:pipeline}
\end{figure}

Figure~\ref{fig:pipeline} depicts the architecture of \tool{}, a modular multi-agent system designed to detect malicious Python packages in the PyPI ecosystem. The system is implemented using the CrewAI framework~\citep{taulli2025crewai}, which facilitates structured collaboration among autonomous agents powered by LLMs. Each agent is instantiated with a distinct semantic responsibility and operates through role-specific prompts, enabling distributed reasoning across the detection pipeline. 

In the current implementation, \tool{} integrates four agents: a package harvester, a source extractor, a code classifier, and a verdict aggregator. All general-purpose agents (\ie \FA, \EA, and \VA) are instantiated using the LLaMA-3 model with prompt-based configuration. The core analytical component, \CA, is powered by a fine-tuned version of CodeBERT, adapted for binary classification of Python source files. This hybrid setup enables both general language-based reasoning and task-specific source code analysis.

\subsection{Agent Responsibilities}

\FA initiates the detection pipeline by retrieving the source code of a given PyPI package. 
It resolves the package name, determines the latest stable version, validates the archive format, and provides a standardized output for downstream analysis. Although this task may appear straightforward, the use of an LLM adds robustness by enabling the agent to interpret user queries in natural language, correct typographical errors, and disambiguate version constraints. This ensures that the system can reliably acquire the intended package even in cases where inputs are noisy or incomplete, thereby reducing the likelihood of upstream failures. The actual download is performed by a lightweight Python utility that interacts with the PyPI JSON API and the \texttt{pip download} command. CrewAI orchestrates this interaction by invoking the script as a subprocess, using the resolved package name and version determined by the Fetcher Agent. This separation ensures that the agent focuses on semantic reasoning and query resolution, while download operations remain deterministic and auditable.

\mybox{\textbf{\small{\FA Prompt}}}{gray!10}{gray!10}{\small{You are a package retriever. You will receive a query for a Python package. Your job is to: \\
		- interpret the package name and version, even if there are typographical errors or incomplete information, \\
		- resolve the latest stable version if no version is provided, \\
		- verify that the package exists in the PyPI repository, \\
		- return a standardized output with the package name, resolved version, and download URL. \\
		
		Output strictly in JSON with the following keys: \\
		"name": resolved package name \\
		"version": resolved version string \\
		"url": download URL of the package source archive \\
		"note": brief comment if disambiguation or correction was applied. \\
}}

\EA receives the downloaded archive and extracts its contents. Its responsibility is to identify relevant \texttt{.py} files while excluding documentation, test cases, configuration files, and other non-code artifacts. Employing an LLM at this stage enables semantic filtering, allowing the system to reason about file relevance based on naming conventions, directory structure, and textual content. These capabilities extend beyond simple rule-based heuristics, helping ensure that downstream analysis focuses on executable code paths where malicious logic is most likely to reside.

\mybox{\textbf{\small{\EA Prompt}}}{gray!10}{gray!10}{\small{You are a file extractor. You will receive a Python package archive with multiple files. Your job is to: \\
		- extract the archive and list all files, \\
		- identify and select only relevant \texttt{.py} source files, \\
		- exclude documentation, test cases, configuration files, or other non-code artifacts, \\
		- return the list of selected files for downstream analysis. \\
		
		Output strictly in JSON with the following keys: \\
		"selected\_files": list of relevant \texttt{.py} file paths \\
		"excluded\_files": list of files skipped with reasons (e.g., test, doc, config). \\
}}

\CA performs the central classification task. Each \texttt{.py} file identified by \EA is analyzed by this agent using a fine-tuned CodeBERT model, which predicts whether the file exhibits characteristics of malicious behavior. The model was fine-tuned using a curated dataset of 6,000 Python files, evenly split between malicious and benign classes. Malicious examples capture a variety of behaviors such as obfuscated payloads, encoded execution strings (e.g., base64), dynamic imports, unauthorized system access via \texttt{subprocess} or \texttt{os}, and outbound network connections. Benign files were drawn from widely used, production-grade PyPI packages with no known security issues, representing standard development practices. The fine-tuning process employed a binary cross-entropy loss function, a learning rate of $2\times 10^{-5}$, a batch size of 16, and four training epochs.

\CA operates at the file level. It receives the raw content of a single \texttt{.py} file and feeds it into the CodeBERT classifier, which returns a binary label: \texttt{malicious} or \texttt{benign}. \revised{CodeBERT is adopted as the classifier because it is a transformer encoder pretrained on a large corpus of source code, and it has demonstrated strong performance in code classification and vulnerability detection tasks, making it well-suited to modeling the semantic and obfuscation patterns present in malicious Python files.} While CodeBERT does not use natural language prompts, the resulting prediction is passed to the \VA, which uses a LLaMA-3 agent to generate a brief natural language rationale. This supports interpretability and provides traceable, human-readable explanations alongside each classification. \revised{The model is trained once in a supervised setting on this curated dataset, and the resulting checkpoint is then used unchanged as the classifier component in all LAMPS experiments.} 

\VA aggregates the per-file predictions from \CA and formulates a package-level verdict. It applies a conservative policy: if any source file within the package is classified as malicious, the entire package is flagged as malicious. This strategy prioritizes recall and reduces the risk of overlooking threats in multi-file packages where only a single module may contain harmful logic. Beyond issuing a binary verdict, \VA synthesizes a natural language justification that references the specific files and behavioral patterns responsible for the decision. An LLaMA-3 instance generates this explanation, prompted with the classification results, enabling transparent, reproducible, and auditable communication of outcomes to end users.

\mybox{\textbf{\small{\VA Prompt}}}{gray!10}{gray!10}{\small{You are a decision aggregator. You will receive file-level
		classifications from other agents. If any file is malicious,
		the package must be classified as MALICIOUS. Otherwise, classify
		it as BENIGN. Provide a short natural language justification
		summarizing the reasoning.
}}

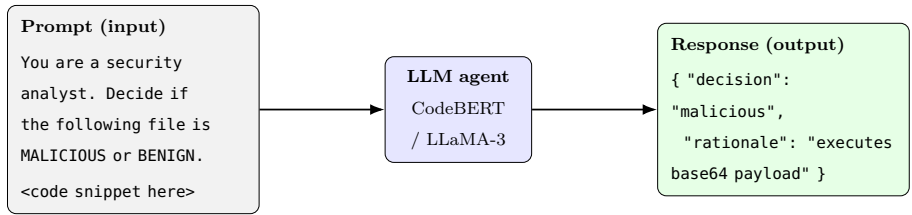
\begin{figure}[h!]
	\centering
	\footnotesize
	\resizebox{\linewidth}{!}{%
		\begin{tikzpicture}[
			node distance=2.0cm,
			>={Latex},
			box/.style={
				draw, rounded corners, align=left, inner sep=6pt,
				minimum height=1.4cm, fill=#1!10
			}
			]
			
			\node[box=gray, text width=0.30\linewidth] (prompt) {
				\textbf{Prompt (input)}\\[2pt]
				{\ttfamily You are a security analyst. Decide if the following file is MALICIOUS or BENIGN.}\\[2pt]
				{\ttfamily <code snippet here>}
			};
			
			\node[box=blue, right=of prompt, text width=0.16\linewidth, align=center, minimum height=1.0cm] (llm) {
				\textbf{LLM agent}\\
				CodeBERT / LLaMA-3
			};
			
			\node[box=green, right=of llm, text width=0.30\linewidth] (resp) {
				\textbf{Response (output)}\\[2pt]
				{\ttfamily \{ "decision": "malicious",}\\
				{\ttfamily \ \ "rationale": "executes base64 payload" \}}
			};
			
			\draw[->, thick] (prompt) -- (llm);
			\draw[->, thick] (llm) -- (resp);
			
		\end{tikzpicture}%
	}
	\caption{Figure 3: Illustrative flow of how a single file-level prediction is processed in LAMPS. The \CA (using CodeBERT) produces a binary label, and the \VA (using LLaMA-3) generates a human-readable rationale in JSON format. 
	}
	\label{fig:promptflow}
\end{figure}

\subsection{Inter-Agent Collaboration}

Agent interactions are coordinated by the CrewAI framework, which facilitates prompt-based communication and contextual handoff. Each agent is instantiated independently but receives contextual input derived from the outputs of upstream components. For example, \EA{}’s execution is conditioned on the archive retrieved by \FA, and \CA{}’s classification prompts are dynamically generated based on the file list produced by \EA. This design supports modular reasoning, allowing each agent to be developed, tested, or replaced independently without compromising the overall system integrity.

An illustrative execution trace is as follows: \FA retrieves a PyPI archive and passes it to \EA, which extracts the relevant \texttt{.py} files. \CA analyzes each file and identifies one as malicious, citing evidence such as base64-decoded subprocess execution. \VA, applying its aggregation policy, marks the entire package as malicious and generates a summary explanation based on the classified file. Each step in this process is logged and can be audited for reproducibility and debugging. 
Figure \ref{fig:promptflow} illustrates the localized interaction between the classifier and verdict agents for a single Python file. Although CodeBERT does not use prompts, we include this prompt metaphorically to show the information passed to the \VA, which wraps the classification result in an interpretable rationale.

\subsection{Threat Model and Detection Scope}

The detection goal of \tool{} is to identify Python packages distributed via PyPI that contain files exhibiting behavior associated with malicious intent. These behaviors are learned from labeled examples during supervised training and are modeled at the source-code level. The system assumes that an attacker may publish a package that appears legitimate in structure and metadata but includes one or more files that attempt to perform harmful actions when executed.

In our threat model, the adversary has complete control over the contents of the published package archive, including its source code and setup scripts. Their objective is to embed harmful logic within these files, such as remote command execution, payload download, or unauthorized access to the local file system. The attacker may also attempt to evade detection through tactics such as obfuscation, encoding, or the use of dynamic language features.

\tool{} makes no assumptions about the package’s runtime behavior or execution environment. Rather, it relies on a classifier trained to detect source-level characteristics of malicious behavior, as reflected in the training data. During training, examples included code patterns such as:
\begin{itemize}
	\item Execution of system commands through APIs like \texttt{subprocess.Popen} or \texttt{os.system}.
	\item Encoded payloads using \texttt{base64} followed by decoding and execution.
	\item Network calls using libraries like \texttt{socket}, \texttt{requests}, or \texttt{urllib}.
	\item Attempts to manipulate files or directories without user awareness.
\end{itemize}

These examples were used to fine-tune a CodeBERT-based binary classifier, which generalizes to unseen files by recognizing similar behavioral signals. The classifier does not perform explicit static analysis, dynamic analysis, or vulnerability scanning; it operates entirely based on patterns learned from labeled Python files.

As such, the scope of detection is limited to code-level indicators of malicious intent present in the analyzed files. Behaviors that depend on runtime context, dynamic dependency resolution, or cross-file logic beyond the accessible file content may remain undetected. Still, \tool{} is capable of identifying a broad class of attacks where malicious logic is expressed explicitly or indirectly in the source code, as demonstrated in our evaluation.

%
%
%

\section{Evaluation}
\label{sec:poc}

This section presents the empirical study conducted on real-world datasets to evaluate the performance of \tool, as well as to compare it with state-of-the-art approaches.

\subsection{Research Questions}

In this article, we conducted a series of experiments to study the proposed approach, answering the following research questions (RQs):

\begin{itemize}
	\item \rqone~\MP~\citep{10298315} is a baseline approach using clustering techniques to group 
	PyPI packages to identify outliers. \revised{Meanwhile, TF-IDF-based stacking~\citep{samaana2025machine} works on top of machine learning and static analysis to examine the package's metadata, code, files, and textual characteristics to detect malicious packages. We experimented to see how well \tool fares against \MP and TF-IDF-based stacking on a large dataset.}
	
	\item \rqtwo~A very recent paper \citep{ibiyo2025detectingmalicioussourcecode} employed RAG 
	as a means to improve the prediction performance. This RQ compares \tool with all the RAG-based 
	configurations introduced in the paper. 
	
	\item \rqthree~We investigate whether the coordinated use of different LLM-based agents is truly worthwhile, given that a single LLM might be sufficient to make accurate predictions. 
\end{itemize}


\subsection{Datasets}

\begin{figure}[h!]
	\centering
	\includegraphics[width=0.95\linewidth]{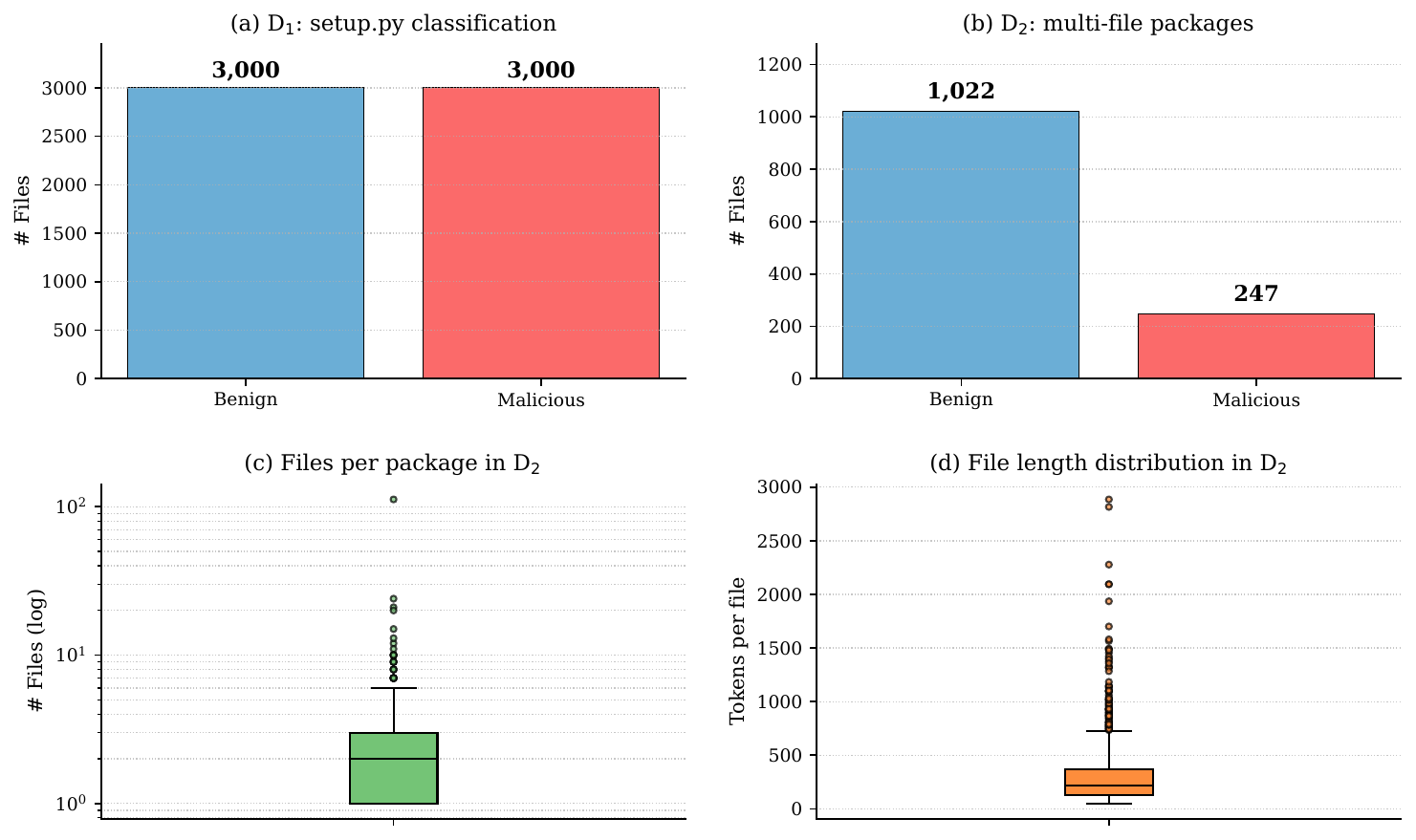}
	\caption{Overview of the datasets. \textbf{(a)} D$_1$: balanced distribution of benign and malicious \texttt{setup.py} files. 
		\textbf{(b)} D$_2$: imbalanced distribution of benign and malicious files across full packages. 
		\textbf{(c)} Distribution of files per package in D$_2$ (log scale), showing most packages are small with a long tail of larger ones. 
		\textbf{(d)} Distribution of file lengths in D$_2$ (tokens), illustrating variability and heavy tails.}
	\label{fig:dataset-fig}
\end{figure}

We evaluated the proposed approach on two datasets, denoted as D$_1$ and D$_2$, which differ in terms of structure, scope, and experimental purpose. Numerical details are reported in Table~\ref{tab:dataset-stats}. Their class balance and structural properties are visualized in Figure~\ref{fig:dataset-fig}.


\begin{table}[!h]
	\centering
	\footnotesize
	\caption{Summary statistics of D$_1$ and D$_2$, reporting number of files, class distribution, and package counts.}
	\label{tab:dataset-stats}
	\begin{tabular}{|l|c|c|c|c|}
		\hline
		\textbf{Dataset} & \textbf{Files} & \textbf{Malicious} & \textbf{Benign} & \textbf{Total Packages} \\
		\hline
		D$_1$ (setup.py) & 6,000 & 3,000 & 3,000 & 6,000 \\
		D$_2$ (multi-file) & 1,296 & 274 & 1,022 & 507 \\
		\hline
	\end{tabular}
\end{table}

\subsubsection{D$_1$: setup.py classification} This dataset comprises 6,000 \texttt{setup.py} files, evenly split between 3,000 malicious and 3,000 benign samples. The malicious samples were collected from a curated dataset of real-world threats in PyPI packages \citep{guo2023empirical}. These samples exhibit behaviors such as base64-encoded payloads, hidden subprocess execution, and system-level command injection via installation scripts. They represent typical installer-time attacks where malicious logic is embedded directly in the setup routine and executed as soon as the package is installed. Figure~\ref{fig:dataset-fig}.a shows the balanced distribution of malicious and benign samples in D$_1$. \revised{Though this one-to-one ratio does not reflect the highly skewed prevalence of malicious packages in the wild, it is intentionally adopted for D$_1$ to enable controlled classifier training, while the subsequent evaluation dataset preserves the real-world imbalance.}

The benign samples were manually collected from popular, production-grade Python packages on PyPI with high download volumes and stable maintenance histories. To minimise the risk of false negatives in the benign class, we ensured that the selected packages were not flagged in vulnerability databases, had not been reported in security advisories, and were not associated with any known security incidents at the time of collection. In addition, each benign sample was verified to represent legitimate usage patterns in installation scripts, for example, specifying metadata, listing dependencies, or defining entry points, without performing obfuscated system calls. 

This balanced dataset is used to compare the classification performance of \tool against \MP, which also operates at the level of \texttt{setup.py} files. The balanced design of D$_1$ provides a controlled setting to measure classification accuracy without confounding factors such as class imbalance or multi-file complexity. As such, it serves as a baseline scenario to evaluate whether the semantic modeling of installer scripts can outperform clustering-based detection of outlier packages.


\subsubsection{D$_2$: Full-package multi-file classification} This dataset was constructed using labeled samples from a prior study by Ibiyo \etal \citep{ibiyo2025detectingmalicioussourcecode}. It includes 274 malicious and 566 benign Python files extracted from complete package archives, thereby capturing realistic multi-file structures as they appear in practice. These malicious files include examples where harmful logic is distributed across modules rather than confined to the installer, for instance, through helper utilities or obfuscated imports that activate at runtime.

To increase the realism and diversity of the dataset, we expanded the benign set by incorporating 456 additional PyPI packages from the most downloaded PyPI packages. 
These packages were selected based on the presence of trusted maintainers, active release histories, and the absence of security flags in public advisories or vulnerability databases. Each new package was manually inspected to confirm that its source files did not contain obvious indicators of malicious behavior, such as encoded payloads, hidden subprocesses, or unauthorized file operations. This careful validation process ensures that benign samples reflect legitimate development practices while maintaining a high level of confidence in their labels.

As a result, D$_2$ consists of 1,296 Python files spanning 507 distinct packages. The dataset exhibits a natural class imbalance, with a higher proportion of benign files, which mirrors real-world conditions where malicious packages are rare compared to legitimate ones. Unlike D$_1$, which contains a single installer script per package, D$_2$ includes packages with multiple modules, thereby enabling the evaluation of detection methods in scenarios that require reasoning across files. Panels (b)--(d) of Figure~\ref{fig:dataset-fig} highlight these properties: panel (b) shows the imbalanced class distribution, panel (c) depicts the skewed distribution of files per package (with most packages small but a long tail of larger ones), and panel (d) presents the distribution of file lengths in tokens, again showing a long-tailed pattern. This property makes D$_2$ particularly suitable for assessing the advantages of multi-agent orchestration, since input curation, file-level classification, and conservative aggregation all play a role in package-level detection. Accordingly, we use D$_1$ for RQ1 and D$_2$ for RQ2 and RQ3, allowing us to contrast performance in balanced single-file settings against more realistic, imbalanced, multi-file conditions.


Table~\ref{tab:dataset-stats} summarizes the number of files, class distribution, and package counts in each dataset. The full source data and scripts used for preprocessing are available in our replication package~\citep{lampsjss2025}.

\subsection{Configurations}

The prototype has been implemented in Python using CrewAI for orchestration and Hugging Face Transformers for model hosting. All experiments were executed locally on an Apple MacBook Pro with an M3 processor through the PyTorch MPS backend. In our configuration, \FA, \EA, and \VA are instantiated with LLaMA-3 via role-specific prompts, while \CA uses a fine-tuned CodeBERT checkpoint for binary file-level prediction with package-level aggregation. Unless otherwise noted, training and inference ran on the same machine.

The single-agent baselines operate at the package level. The model receives multiple files concatenated into a single prompt. Two regimes were tested: SA--Concat (all files concatenated until the context window is reached) and SA--TopK (the three longest or most relevant files). The prompt used to guide single agents is shown below.

\mybox{\textbf{\small{Single-Agent Prompt}}}{gray!10}{gray!10}{\small{	You are a security auditor. Analyze the following files
		belonging to a single Python package. Based only on the code
		shown, decide whether the package is MALICIOUS or BENIGN.
		Pay particular attention to files such as setup.py scripts,
		payload downloads, subprocess execution, or obfuscated strings.\\
		
		Output strictly in JSON with two keys:
		"decision": either "malicious" or "benign"
		"brief\_rationale": a short justification referring to
		the relevant code patterns.
}}


%

\section{Experimental Results}
\label{sec:ExperimentalResult}

For all comparisons, we use 
the datasets described earlier, with splits defined at the package level whenever multiple files belong to the same package, ensuring that no file from a given package appears in both the training and test partitions. 
We report Accuracy, Precision, Recall, and F1; for imbalanced settings, we additionally report Balanced Accuracy. Statistical significance for paired predictions is assessed with McNemar’s test \citep{pembury2020effective} on identical test instances, consistent with the analysis reported in this section. To mitigate randomness, we repeat each experiment with fixed seeds and report mean values together with the corresponding standard deviations. Library versions are pinned in the released environment file to improve determinism on the MPS backend. The single-agent baseline uses an LLaMA-3 prompt that receives the list of extracted file paths and their concatenated contents, and is instructed to output a package-level verdict with a brief justification. The exact template is included in the replication package. We also record per-package latency and memory usage on the M3 system to report efficiency alongside accuracy.

\subsection{\rqone}

With respect to the motivating example in Listing~\ref{lst:malicious-example2}, \MP cannot recognize 
the silently \shadedtext{base64-encoded} command, failing to detect the hidden malicious intent. This is consistent with a clustering-based mechanism that groups samples by surface-level similarity: obfuscated, single-file installer logic may not exhibit the frequency-based characteristics required to form a clear outlier cluster. In contrast, \tool correctly flags the snippet as malicious thanks to the learned behavioral signals used by the \CA{}. In our configuration, \CA{} has been fine-tuned on labeled files that include paired patterns of encoding followed by execution, which enables it to discriminate obfuscated installer logic from benign configuration statements without relying on handcrafted signatures. The classifier is guided through a fixed prompt that requires a binary label (\emph{malicious} or \emph{benign}) and a short rationale, ensuring consistent outputs across runs. This highlights \tool's ability to identify obfuscated threats capable of bypassing static heuristics.

	\noindent
	$\triangleright$ 
	\revised{\textbf{Example.} A representative malicious sample from D$_1$ is shown in Listing~\ref{lst:malicious-example}. It contains an obfuscated PowerShell payload that runs only when a marker file is missing, making the installer appear benign at first glance.}

\begin{lstlisting}[style=pythonstyle,
	caption={Excerpt from the malicious \texttt{setup.py} of package \texttt{pongreplace-10.4}, showing the obfuscated payload execution.},
	label={lst:malicious-example}]
	import subprocess, os
	if not os.path.exists('tahg'):
	subprocess.Popen(
	'powershell -WindowStyle Hidden -EncodedCommand '
	'cABvAHcAZQByAHMAaABlAGwAbAAgAEkAbgB2AG8AawBlAC0AVwBlA...',
	shell=False,
	creationflags=subprocess.CREATE_NO_WINDOW)
\end{lstlisting}

\revised{
	\noindent
	\MP assigns a benign label to this package because the installer script resembles ordinary configuration code, and the obfuscated command appears as an isolated, rare string that does not form a clear outlier cluster. \tool flags the package as malicious instead. It explains that the conditional check on a marker file, combined with a hidden encoded PowerShell command executed during installation, is inconsistent with legitimate \texttt{setup.py} behaviour. This case illustrates how \tool can correctly classify obfuscated installer time attacks that remain unnoticed by clustering-based analysis.
}

\begin{table}[h]
	\centering
	\scriptsize
	\caption{Comparison between \MP, supervised TF-IDF based stacking, and \tool.}
	\label{tab:classification-result}
	\begin{tabular}{|l|r|r|r|r|} \hline
		\textbf{Approach} & \textbf{Accuracy (\%)}  & \textbf{P} & \textbf{R} & \textbf{F1} \\ \hline
		\MP (Unsupervised baseline)  & 95.86 & 0.963  & 0.955 & 0.957 \\   \hline
		\revised{TF-IDF + Stacking (Supervised baseline)} & \revised{88.45} & \revised{0.874} & \revised{0.899} & \revised{0.887} \\ \hline
		\tool & \textbf{97.77} & \textbf{0.977} & \textbf{0.977} & \textbf{0.977}  \\ \hline
		\revised{Performance Gain vs \MP} 
		& \revised{\color{darkgreen}$\uparrow$2.0\%} 
		& \revised{\color{darkgreen}$\uparrow$1.4\%}  
		& \revised{\color{darkgreen}$\uparrow$2.32\%}  
		& \revised{\color{darkgreen}$\uparrow$2.0\%} \\ \hline
		\revised{Performance Gain vs TF-IDF + Stacking} 
		& \revised{\color{darkgreen}$\uparrow$10.5\%} 
		& \revised{\color{darkgreen}$\uparrow$11.8\%}  
		& \revised{\color{darkgreen}$\uparrow$8.7\%}  
		& \revised{\color{darkgreen}$\uparrow$10.1\%} \\ \hline
	\end{tabular}
\end{table}

\smallskip
\noindent
$\triangleright$ \revised{\textbf{Comparison with unsupervised \MP.} \MP is an unsupervised clustering method, while \tool builds on a supervised classifier. Nevertheless, the comparison remains meaningful because both techniques aim to solve the same problem, and both operate on installer-level information extracted from PyPI packages. To address the methodological gap fairly, we ensured that the evaluation protocol aligned with the assumptions underlying each technique. More concretely, \tool trains its classifier on the D$_1$ training set and is evaluated on a held-out test split. In contrast, \MP, which does not require training, is executed only on the same test packages. \MP produces anomaly scores on this test set, which we threshold into malicious or benign labels, and then compare them with the ground-truth labels. In this way, both methods are evaluated on exactly the same test files and labels, ensuring a controlled comparison of detection effectiveness.}
The prediction results for \tool and \MP 
on \textbf{D$_1$} are shown in Table \ref{tab:classification-result}, which confirms that \MP obtains encouraging performance as claimed by the authors~\citep{10298315}, \ie it gets 95.86\% as Accuracy, 0.963 as Precision, 0.955 as Recall, and 0.957 F1-score. Still, \tool is better as it gains more accurate predictions with respect to all the metrics, \ie, Accuracy is 97.77\%, and all Precision, Recall, and F1-score are equal to 0.977.\footnote{These scores are equal thanks to the balanced dataset (\textbf{D$_1$}).}

\begin{figure}[h!]
	\centering
	\includegraphics[width=0.99\linewidth]{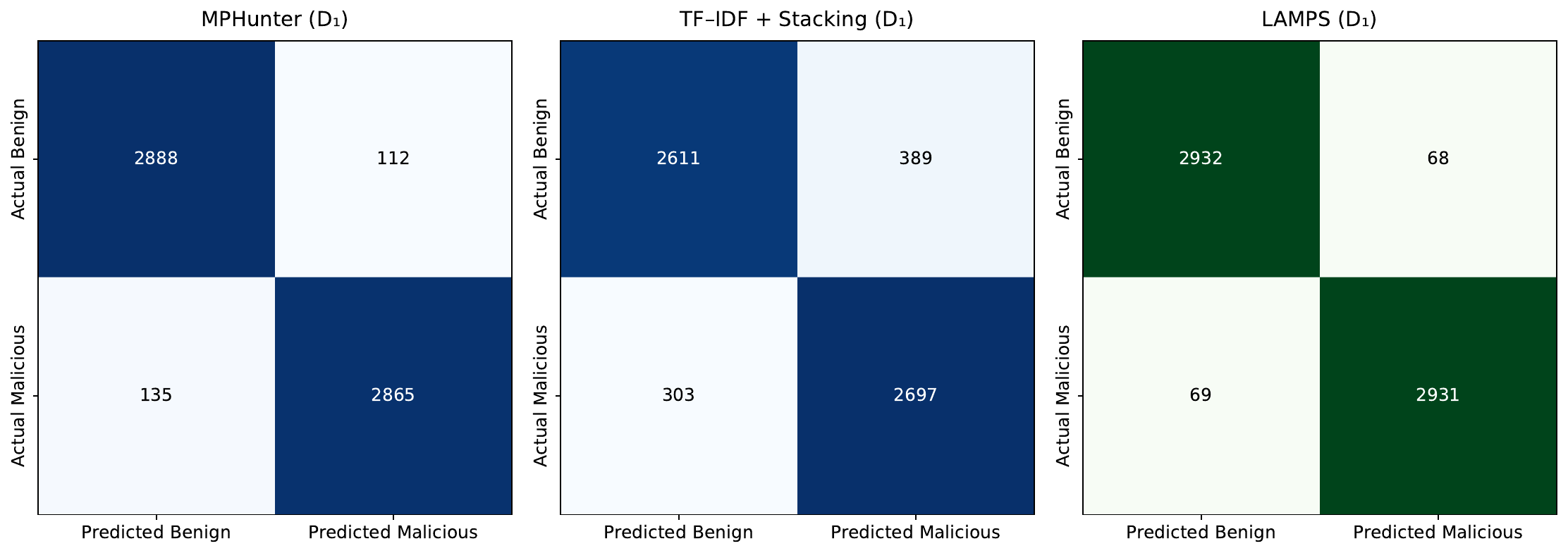}
	\revised{\caption{Confusion matrices for \MP, TF-IDF + Stacking and \tool on D$_1$. Compared to \MP, and TF- IDF +Stacking baselines, \tool significantly reduces misclassifications, achieving a better balance between benign and malicious detection.}}
	\label{fig:RQ1ConfMat-fig}
\end{figure}

\smallskip
\noindent
$\triangleright$ \revised{\textbf{Comparison with supervised TF-IDF Stacking model.} Apart from \MP, to provide a supervised point of comparison beyond clustering-based detection, we compared \tool with a TF-IDF baseline implemented by recent work on malicious package identification~\citep{samaana2025machine}. The baseline trains a stacking ensemble on D$_1$ and evaluates it on unseen real-world packages drawn from our broader corpus. This setting reflects a realistic deployment scenario and allows a direct comparison between signature-leaning lexical models and the semantic reasoning performed by \tool. As shown in  Table \ref{tab:classification-result}, the supervised TF-IDF based stacking model performs reasonably well on unseen real-world files, achieving an F1-score of 0.887, nevertheless it still underperforms \tool, highlighting the benefits of semantic reasoning across agents rather than relying on lexical similarity alone.} 
The evaluation follows the protocol described at the beginning of this section, with splits at the package level to avoid leakage. Results are averaged over fixed seeds; standard deviations and exact prompt templates are included in the replication package for reproducibility. The confusion matrices in Figure~\ref{fig:RQ1ConfMat-fig} further illustrate this improvement, showing that \tool reduces both false positives and false negatives compared to \MP.

\smallskip
\noindent
$\triangleright$ \textbf{Statistical Analysis.} 
We ran McNemar's test \citep{pembury2020effective} on the predictions of \MP and \tool over the same test set of 6,000 \texttt{setup.py} files to compare paired outcomes, focusing on cases where the two models disagree. The resulting p-value is $9.63 \times 10^{-18}$, indicating that the improvement by \tool compared to \MP is statistically significant at a very high confidence level, as $p \ll 0.05$. To complement the significance test, we also computed the effect size to assess the magnitude of the difference and monitored the stability of the results across repeated runs. Effect sizes, per-seed variability, and the corresponding confidence intervals are provided in the replication package together with the exact random seeds, environment specification, and full prompt text. This ensures that the reported improvement is not only statistically significant but also practically meaningful and robust to sources of randomness inherent in training and evaluation.
\revised{We also compared \tool with the supervised TF-IDF + Stacking baseline using McNemar's test on the same paired predictions in D$_1$. The resulting p-value is below $10^{-20}$, which satisfies $p \ll 0.05$ and indicates that the improvement of \tool over the supervised baseline is statistically significant. This confirms that the observed gains are consistent across both unsupervised and supervised baselines.}

\vspace{.1cm}
\begin{shadedbox}
	\small{\textbf{Answer to $RQ_1$.} 
		\revised{Compared to the baselines \MP and TF-IDF based stacking model, \tool achieves a higher performance by all the considered metrics. Moreover, the observed improvement is statistically significant.}}
\end{shadedbox}

\subsection{\rqtwo} 

Recently, in the RAG-based approach by \citep{ibiyo2025detectingmalicioussourcecode}, external knowledge such as YARA rules, GitHub Security Advisories, and malicious \texttt{setup.py} files were embedded using OpenAI’s \texttt{text-embedding-ada-002} model and stored in a vector database. For each input, relevant documents were retrieved based on semantic similarity and concatenated with the code or its abstract syntax tree (AST). This combined input was then passed to a language model, including LLaMA-3.1-8B, to produce a classification. While the goal was to improve detection by providing contextual information, the results showed that RAG-based methods performed worse than zero-shot and fine-tuned baselines. Their experiments were conducted on the same D$_2$ dataset, allowing a direct comparison for our experiments. 

In this RQ, we compare \tool with all these settings on the same dataset to assess whether distributed multi-agent reasoning can offer advantages over retrieval-augmented generation. Consistent with their setup, we reuse D$_2$ exactly as described earlier and replicate each configuration following the authors' stated retrieval sources and parameters. To ensure reproducibility, we employ the same package-level splits and fixed seeds, and we also preserve the original prompt formulations. For \tool, \CA uses a file-level prompt requiring a binary decision with a brief rationale, while the single-agent RAG configurations are instructed to return package-level verdicts. The exact templates are reported 
in the replication package.

\smallskip
\noindent
$\triangleright$ 
\revised{
	\noindent\textbf{Example.} A representative D$_2$ package illustrates how \tool differs from RAG-based methods. The \texttt{setup.py} file of package \texttt{streamsyncer} only declares metadata and imports a helper module, so its content closely resembles a benign installer, while the actual payload is implemented in a separate helper file. An anonymised version of this pattern is shown in Listing~\ref{lst:rq2-example}, which is derived from a real malicious project in our corpus.
}

\begin{lstlisting}[style=pythonstyle,
	caption={Example from D$_2$ where the installer appears benign but the payload is activated from a helper module.},
	label={lst:rq2-example}]
	# setup.py (benign-looking installer)
	from setuptools import setup
	import helper_trigger
	
	setup(
	name="streamsyncer",
	version="2.9",
	packages=["streamsyncer"],
	)
	
	# helper_trigger.py (malicious helper)
	import base64, subprocess, tempfile
	
	blob = "cABvAHcAZQByAHMAaABlAGwAbAAgAEkAbgB2AG8AawBl..."
	tmp = tempfile.NamedTemporaryFile(delete=False)
	tmp.write(base64.b64decode(blob)); tmp.close()
	subprocess.Popen(tmp.name)
\end{lstlisting}

\revised{
	\noindent
	RAG-based configurations frequently predict a benign label for this package, because retrieval is driven by the surface content of \texttt{setup.py}, which does not contain explicit indicators of malicious behaviour. \tool instead harvests all project files, lets the Extractor and Classifier Agents jointly analyse both \texttt{setup.py} and \texttt{helper\_trigger.py}, and the Verdict Synthesizer Agent correctly marks the package as malicious based on the cross file execution chain. Similar patterns appear in multiple D$_2$ samples and help explain the higher recall of \tool on malicious packages compared to the RAG-based baselines.
}

\begin{figure}[t]
	\centering
	\includegraphics[width=\linewidth]{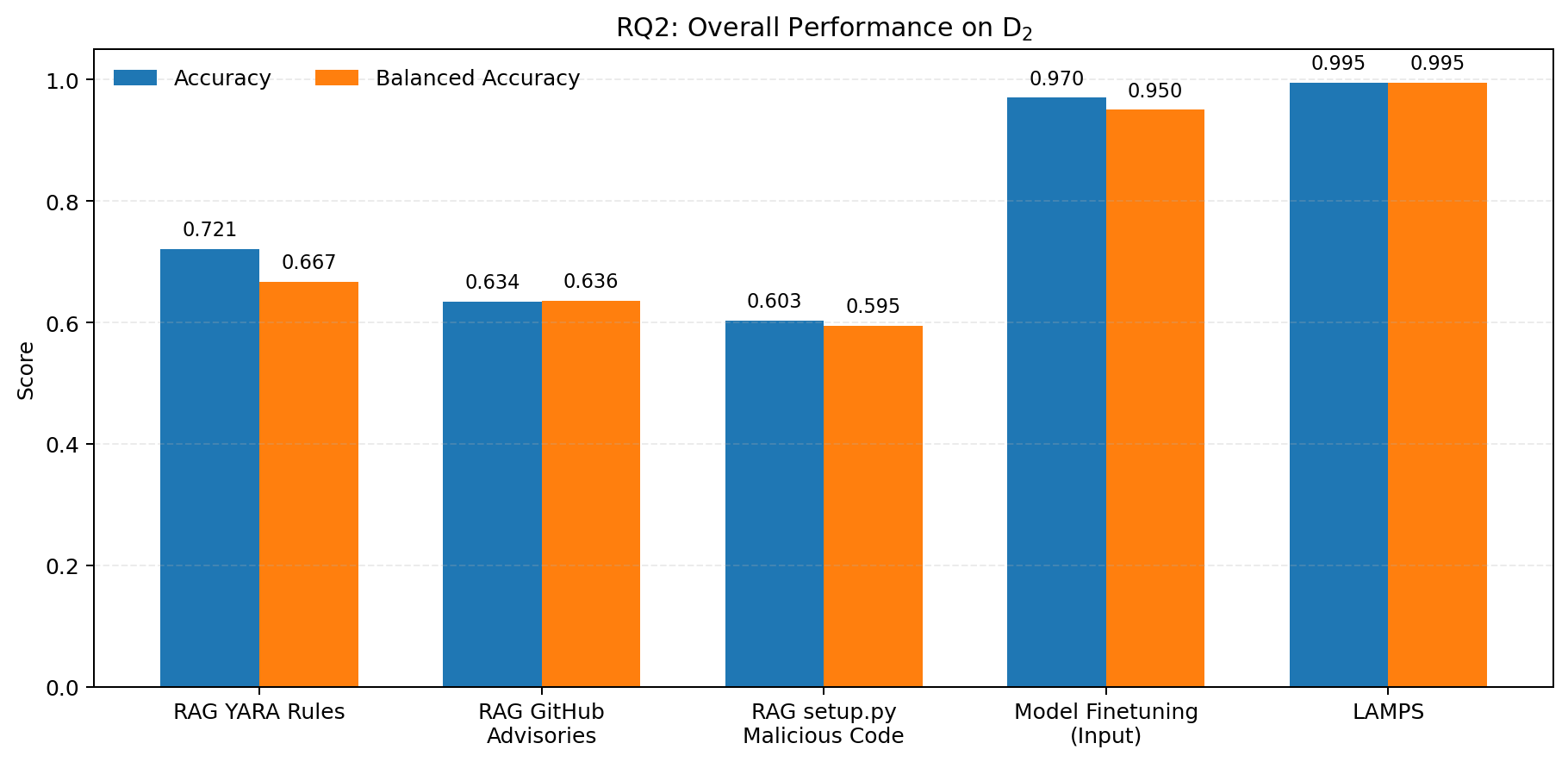}
	\caption{RQ$_2$ overall performance on D$_2$: Accuracy and Balanced Accuracy for each method.}
	\label{fig:rq2-overall}
\end{figure}

\begin{figure}[t]
	\centering
	\includegraphics[width=\linewidth]{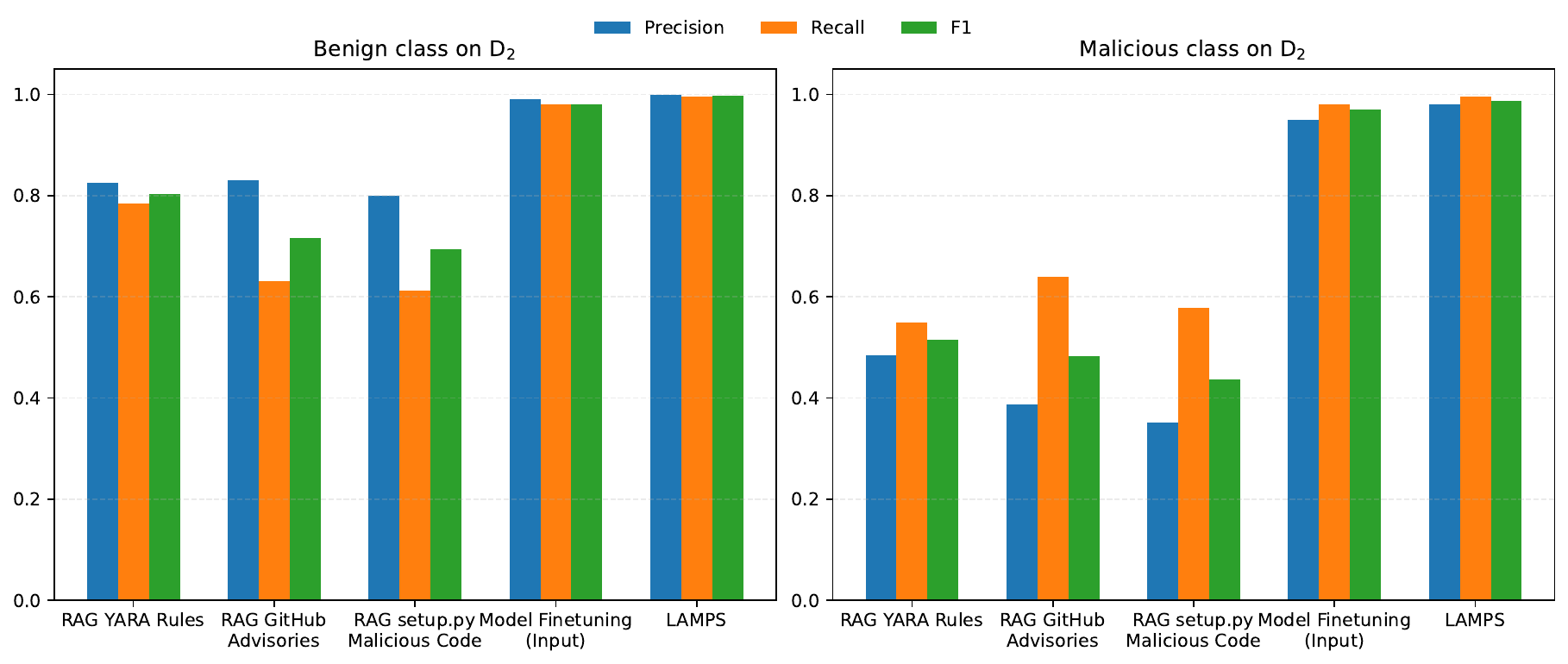}
	\caption{RQ$_2$ per-class performance on D$_2$. Precision, Recall, and F1 for \emph{Benign} (left) and \emph{Malicious} (right) classes. \tool consistently improves both classes, whereas RAG-based methods struggle particularly on the malicious class.}
	\label{fig:rq2-per-class}
\end{figure}

Figure~\ref{fig:rq2-overall} shows the overall results in terms of accuracy and balanced accuracy, where \tool clearly outperforms all RAG-based methods, including the fine-tuned variant. Figure~\ref{fig:rq2-per-class} further breaks down the evaluation by class, illustrating that RAG approaches achieve relatively high scores on benign files but suffer severe drops on malicious ones. In contrast, our proposed approach maintains high precision and recall for both classes, highlighting its robustness in identifying rare malicious packages. 

Table~\ref{tab:RAGbasedAndLAMPS} reports the detailed comparison between \tool and the RAG-based baselines. \tool consistently exceeds the three retrieval-only configurations (YARA Rules \citep{YARAForge2024}, GitHub Security Advisories, and raw \texttt{setup.py} snippets) across all metrics. For instance, the highest precision among those variants is 0.825 (YARA), whereas \tool achieves 0.999. The same pattern holds for recall and F1 on both benign and malicious classes, and the overall superiority is reflected in accuracy and balanced accuracy. When the baseline is enhanced with model fine-tuning, 
its performance improves substantially (e.g., precision 0.970) but still remains below \tool on every reported metric. These gains are robust to seed choice and are averaged across repeated runs, with variance across seeds reported in the replication package.

\begin{table}[t]
	\centering
	\caption{RQ$_2$: Comparison of \tool with RAG-based Approaches \citep{ibiyo2025detectingmalicioussourcecode}.}
	\label{tab:RAGbasedAndLAMPS}
	\setlength{\tabcolsep}{3.5pt}
	\renewcommand{\arraystretch}{1.6}
	\resizebox{\linewidth}{!}{%
		\begin{tabular}{lcccccccccc}
			\hline
			\textbf{Approaches} &
			\multicolumn{4}{c}{\textbf{Benign}} &
			\multicolumn{4}{c}{\textbf{Malicious}} &
			\textbf{Accuracy} &
			\textbf{Balanced Acc.} \\
			\hline
			& \textbf{P} & \textbf{R} & \textbf{F1} & \textbf{Support}
			& \textbf{P} & \textbf{R} & \textbf{F1} & \textbf{Support}
			& & \\
			\hline
			RAG YARA Rules           & 0.825 & 0.784 & 0.804 & 750
			& 0.484 & 0.550 & 0.515 & 276
			& 0.721 & 0.667 \\
			RAG GitHub Advisories    & 0.831 & 0.631 & 0.717 & 747
			& 0.387 & 0.640 & 0.483 & 270
			& 0.634 & 0.636 \\
			RAG \texttt{setup.py} Malicious Code
			& 0.800 & 0.613 & 0.694 & 752
			& 0.351 & 0.578 & 0.437 & 273
			& 0.603 & 0.595 \\
			Model Finetuning (Input) & 0.970 & 0.990 & 0.980 & 376
			& 0.980 & 0.950 & 0.970 & 133
			& 0.970 & 0.950 \\
			\rowcolor{green!20}
			\tool                    & \textbf{0.999} & \textbf{0.995} & \textbf{0.997} & 1,022
			& \textbf{0.980} & \textbf{0.996} & \textbf{0.988} & 247
			& \textbf{0.995} & \textbf{0.995} \\
			\hline
	\end{tabular}}
\end{table}

A plausible explanation, consistent with our methodology, is that retrieval alone can dilute or miss file-local signals when installer logic is obfuscated, fragmented across modules, or sparsely expressed. In such cases, retrieving context from external corpora does not necessarily surface the most relevant indicators of malicious intent. By contrast, \tool relies on a fine-tuned file-level classifier that directly models malicious behaviors present in code and complements it with a conservative package-level aggregation strategy. This combination ensures that critical signals are preserved and reduces the likelihood of missed detections in multi-file workflows. Moreover, the modular outputs of \tool enable auditability, as intermediate classifications and rationales can be inspected, which is not the case with monolithic RAG prompts.

\vspace{.1cm}
\begin{shadedbox}
	\small{\textbf{Answer to $RQ_2$.} \tool provides consistent advantages over RAG-based approaches on D$_2$, outperforming all four configurations across evaluation metrics and offering greater stability and interpretability. The improvements derive from file-level specialization and conservative aggregation, which preserve critical signals that retrieval alone may dilute or omit.}
\end{shadedbox}

\subsection{\rqthree} 

This RQ compares \tool with detection engines configured as single-agent baselines, to assess whether task specialization and inter-agent coordination offer measurable advantages over a monolithic prompt. Experiments are conducted on \textbf{D$_2$} under the same protocol used in this section, with package-level splits and repeated runs over fixed seeds to mitigate randomness. The primary single-agent baseline is an LLaMA-3 model operating as a direct package-level classifier: it receives the list of extracted file paths, along with their concatenated contents, as a single prompt and is instructed to output a binary verdict (\emph{malicious} or \emph{benign}) along with a brief justification. To ensure consistency, the prompt was fixed across all runs, and its exact text is provided 
in the replication package \citep{lampsjss2025}.

\begin{figure}[t]
	\centering
	\includegraphics[width=0.9\linewidth]{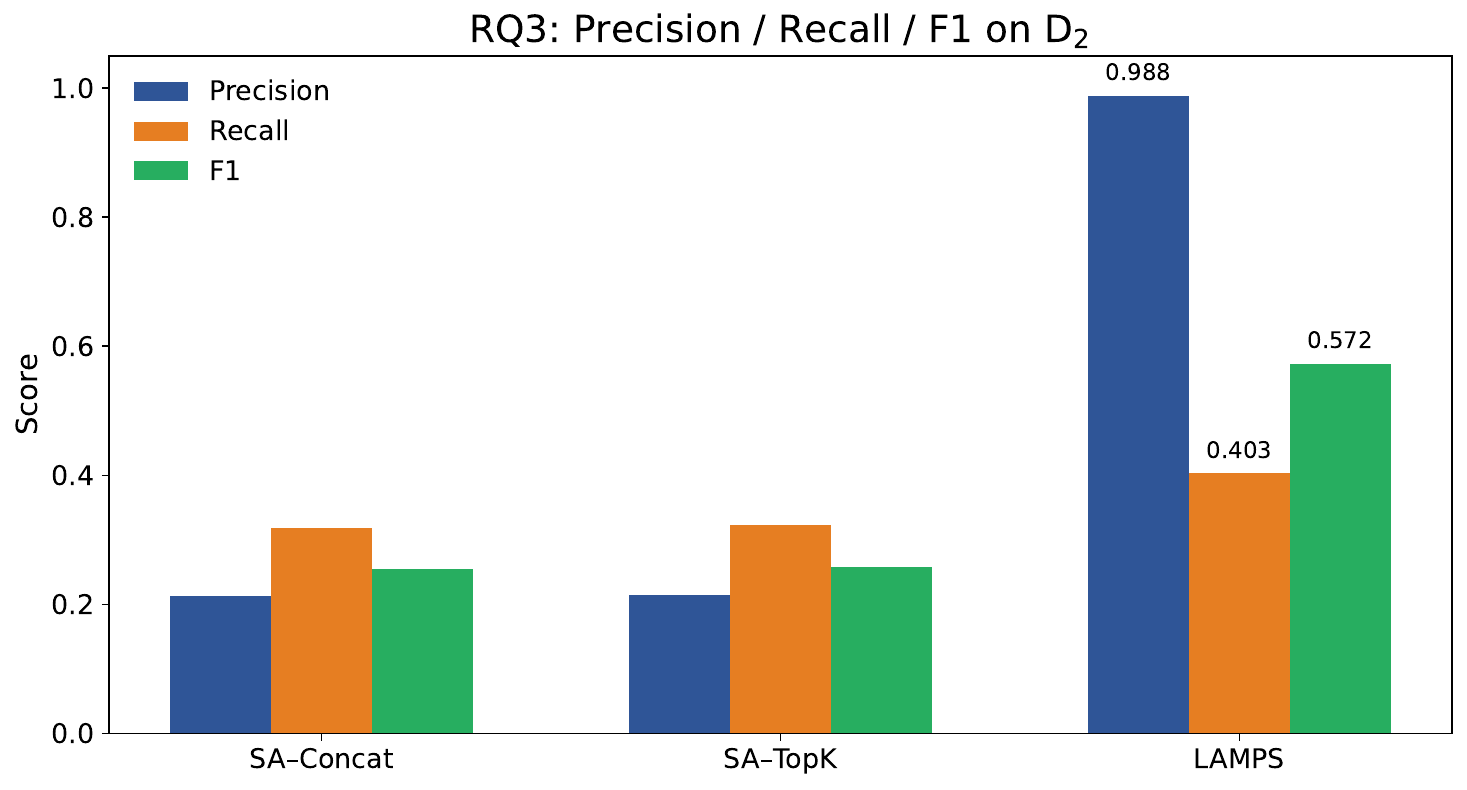}
	\caption{RQ$_3$ precision, recall, and F1 on D$_2$. \tool achieves substantially higher recall while maintaining strong precision and F1 compared to single-agent baselines (SA--Concat and SA--TopK).}
	\label{fig:rq3-metrics}
\end{figure}

Figure~\ref{fig:rq3-metrics} compares precision, recall, and F1 across the two single-agent baselines and \tool. The figure shows that while the single-agent regimes maintain moderate precision, they struggle with recall, missing a substantial portion of malicious files. In contrast, \tool achieves a much stronger balance, sustaining high recall without sacrificing precision, which translates into a higher F1 score overall.

To isolate the contributions of input curation and package-level aggregation, we further evaluate two constrained single-agent regimes. In the first regime (\textbf{SA-Concat}), the agent processes all extracted files in a deterministic order until the model’s context window is reached, at which point the remaining files are truncated. This concatenation-based approach has been considered in prior work on code analysis with LLMs~\citep{choi2023codeprompt,ahmad2021unified}. In the second regime (\textbf{SA-TopK}), the agent processes only the top $K$ files identified by the extractor as most relevant to package functionality and security; we set $K{=}3$ to ensure that typical packages remain within the context window. Similar top-$K$ retrieval heuristics have been explored in LLM-augmented code search and software engineering tasks~\citep{chen2023autoagents}. Both regimes used a fixed prompt instructing the model to produce a binary decision with a brief rationale; the full template is reported in the online appendix. 
Together, these baselines highlight the limitations of a monolithic LLM in multi-file scenarios where input length and ordering can affect detection outcomes.

\begin{table}[t]
	\centering
	\footnotesize
	\caption{RQ$_3$: Comparison of \tool with single-agent baselines on D$_2$ (package level). 
		\tool significantly outperforms both single-agent regimes under McNemar’s test ($p \ll 0.001$).}
	\label{tab:rq3-results}
	\setlength{\tabcolsep}{6pt}
	\renewcommand{\arraystretch}{1.15}
	\begin{tabular}{lccccc}
		\hline
		\textbf{Method} & \textbf{Acc.} & \textbf{Prec.} & \textbf{Rec.} & \textbf{F1} & \textbf{Bal. Acc.} \\
		\hline
		SA--Concat & 0.766 & 0.213 & 0.318 & 0.255 & 0.574 \\
		SA--Top$K$ & 0.766 & 0.215 & 0.323 & 0.258 & 0.576 \\
		\rowcolor{green!20} \tool{} & 0.924 & 0.988 & 0.403 & 0.572 & 0.701 \\
		\hline
	\end{tabular}
\end{table}

For comparison, \tool remains unchanged and continues to classify at the file level using a fine-tuned CodeBERT classifier and to apply a conservative package-level aggregation policy. This modular decomposition ensures that malicious signals present in even a single file are preserved at the package level, rather than being diluted or omitted when inputs exceed the context budget.

As shown in Table~\ref{tab:rq3-results}, \tool achieves substantially stronger package-level detection on D$_2$, with an accuracy of 0.924, precision of 0.988, recall of 0.403, F1 of 0.572, and balanced accuracy of 0.701. In contrast, the single-agent baselines perform markedly worse: SA-Concat attains 0.766 accuracy (precision 0.213, recall 0.318, F1 0.255, balanced accuracy 0.574) and truncates 16.9\% of packages due to context limits; SA-TopK, which limits the input to three files per package, shows similar performance (accuracy 0.766, precision 0.215, recall 0.323, F1 0.258, balanced accuracy 0.576). These results confirm that multi-file signals are diluted or lost when inputs are constrained to fit within a single prompt.

\begin{figure}[h!]
	\centering
	\includegraphics[width=0.9\linewidth]{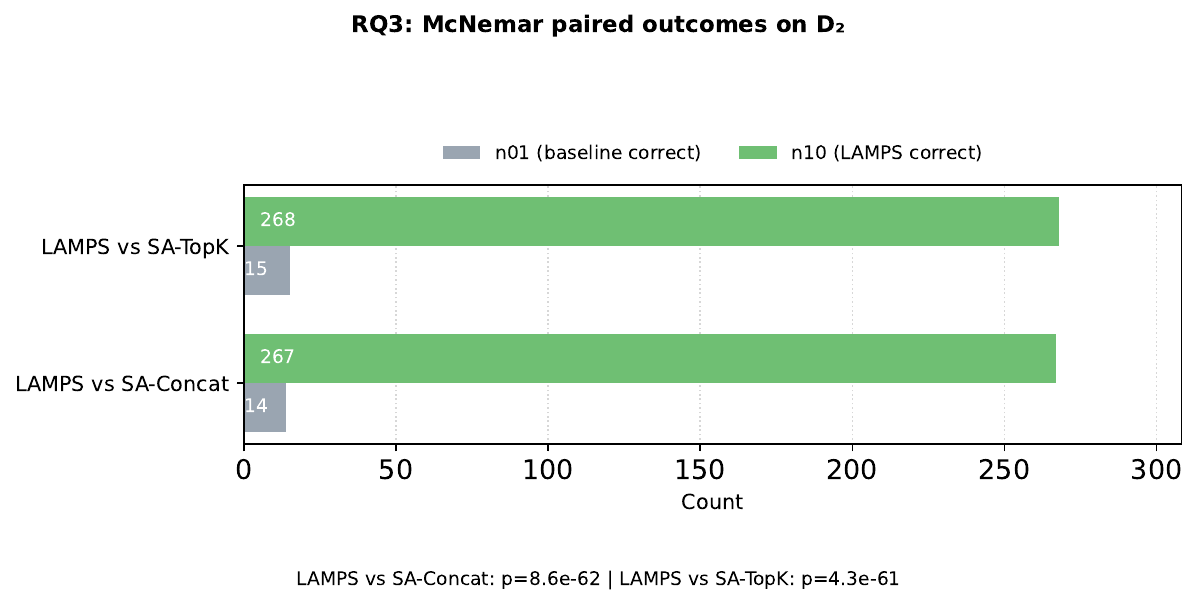}
	\caption{RQ$_3$ McNemar paired outcomes on D$_2$. For each comparison, $n_{01}$ counts packages where the baseline is correct and \tool is wrong, while $n_{10}$ counts the reverse. The strong skew towards $n_{10}$ indicates that \tool significantly outperforms both single-agent regimes ($p \ll 0.001$).}
	\label{fig:rq3-mcnemar}
\end{figure}

Figure~\ref{fig:rq3-mcnemar} complements this analysis by reporting McNemar’s test outcomes, which highlight the large imbalance between $n_{01}$ and $n_{10}$. The skew strongly favors \tool, providing robust statistical evidence that its advantages over single-agent regimes are not due to chance.

\begin{table}[h!]
	\centering
	\footnotesize
	\caption{McNemar’s test results for RQ$_3$ on D$_2$. 
		$n_{01}$ = \#packages where baseline is correct and \tool is wrong; 
		$n_{10}$ = \#packages where \tool is correct and baseline is wrong. 
		All differences are highly significant ($p \ll 0.001$).}
	\label{tab:rq3-mcnemar}
	\setlength{\tabcolsep}{8pt}
	\renewcommand{\arraystretch}{1.15}
	\begin{tabular}{lccc}
		\hline
		\textbf{Comparison} & $n_{01}$ & $n_{10}$ & $p$-value \\
		\hline
		\tool{} vs.\ SA--Concat & 14 & 267 & $8.58{\times}10^{-62}$ \\
		\tool{} vs.\ SA--Top$K$ & 15 & 268 & $4.27{\times}10^{-61}$ \\
		\hline
	\end{tabular}
\end{table}

Paired McNemar’s tests indicate that the differences are highly significant: for SA-Concat, $n_{01}{=}14$, $n_{10}{=}267$, $p{=}8.58{\times}10^{-62}$; for SA-TopK, $n_{01}{=}15$, $n_{10}{=}268$, $p{=}4.27{\times}10^{-61}$ as indicated in Table \ref{tab:rq3-mcnemar}. The conservative aggregation in \tool therefore provides a statistically reliable improvement over both monolithic configurations.

\vspace{.1cm}
\begin{shadedbox}
	\small{\textbf{Answer to $RQ_3$.} The multi-agent design of \tool provides clear advantages over single-agent baselines, as task specialization and explicit aggregation yield more accurate and robust package-level detection than monolithic LLMs affected by context limitations and order sensitivity.}
\end{shadedbox}

\subsection{Discussion}

\subsubsection{Why a multi-agent design?}

Although the \CA performs the discriminative task, the remaining agents are functional components rather than auxiliary glue. The \FA and \EA enforce provenance and input discipline by normalizing archives, removing non-functional material such as tests and documentation, and imposing a deterministic ordering of files. This curation maintains input context limits, preserves file-local signals for the classifier, and prevents accidental leakage across packages when constructing inputs. The \VA turns file-level predictions into an explicit package-level decision under a conservative policy and produces a concise justification that can be audited. These roles improve reproducibility, reduce variance across runs, and isolate failures when a step misbehaves. 

The design therefore supports accuracy, stability, and interpretability in multi-file workflows. In addition, the separation of roles yields measurable efficiency benefits: input curation reduces token volume and context-window pressure, which lowers latency and memory usage on the evaluation hardware. The explicit hand-off between agents establishes verifiable interfaces that can be logged and replayed, enabling consistent replication of results across seeds and environments. The architecture is also extensible: replacing or adapting the fetcher and extractor is sufficient to target other ecosystems without modifying the classifier or the aggregation policy. The verdict stage provides a single, auditable point for calibrating precision–recall trade-offs at the package level. The modularity further facilitates error analysis and ablation: when performance differs across settings, the contribution of curation and aggregation can be quantified independently of the classifier’s discrimination.

\subsubsection{Open Issues}

This work represents an early step toward understanding how task decomposition and inter-agent collaboration can enhance the performance and trustworthiness of LLM-based software analysis. Two directions are critical. First, the design of roles and hand-offs should be formalized so that choices, for example, merging fetching and extraction versus keeping them separate, are guided by measurable criteria. A principled framework would specify the contract of each agent, the information passed between agents, and the expected failure modes, and would be evaluated through ablations that quantify the effect of role changes on accuracy, balanced accuracy, latency, and token usage. Second, the balance between localized reasoning at the file level and aggregation at the package level requires a systematic study to clarify when aggregation is beneficial and when it may obscure relevant signals. This analysis should include sensitivity to aggregation policies and thresholds, robustness to package size and file ordering, and the impact of class imbalance in multi-file settings. 

Explanations produced by the system should be evaluated as first-class outputs and linked to concrete code regions. Beyond qualitative inspection, the fidelity and stability of explanations across seeds and prompt variants should be measured to support interpretability claims with empirical evidence. 
\revised{
	For our work, efficiency metrics were collected on a MacBook Pro M3 Pro using the MPS backend (11-core CPU, 14-core GPU, and 18 GB unified memory). For each agent in \tool, we report average latency per invocation, throughput in processed files per second, and peak resident memory usage, together with prompt and completion token counts for LLM-based components. Measurements are computed on representative samples from the evaluation datasets, using the same configuration as in the main experiments. Table~\ref{tab:efficiency} summarises these results and characterises the computational footprint of \FA, \EA, \CA, and \VA 
	as well as the \MP and fine-tuned LLaMA baselines.
}

\begin{table}[h!]
	\centering
	\setlength{\tabcolsep}{4pt}
	\renewcommand{\arraystretch}{1.1}
	\scriptsize
	\footnotesize
	\caption{\revised{Efficiency metrics on the evaluation hardware (MacBook Pro M3 Pro, MPS backend). Latency and throughput are measured per agent call. Memory is peak resident set size. Token counts refer to average prompt plus completion tokens for LLM-based agents.}}
	\label{tab:efficiency}
	\resizebox{\linewidth}{!}{%
		\begin{tabular}{lcccc}
			\hline
			\textbf{Component} 
			& \textbf{Latency (ms)} 
			& \textbf{Throughput (files/s)} 
			& \textbf{Peak memory (GB)} 
			& \textbf{Tokens / decision} \\
			\hline
			\FA (LLaMA) 
			& 145  
			& 6.8 
			& 5.90 
			& 250 \\
			\EA (LLaMA) 
			& 172  
			& 5.5 
			& 6.00 
			& 320 \\
			\CA (CodeBERT) 
			& 118 
			& 8.4  
			& 1.92 
			& 430 \\
			\VA (LLaMA-3-8B) 
			& 265 
			& 3.1  
			& 6.25 
			& 690 \\
			\MP 
			& 54  
			& 18.5 
			& 0.61 
			& 0 \\
			Fine-tuned LLaMA baseline 
			& 312 
			& 2.6  
			& 6.55 
			& 720 \\
			\hline
		\end{tabular}
	}
\end{table}

\subsubsection{Limitations}

The current prototype targets Python packages from PyPI and relies on a fine-tuned CodeBERT classifier together with three LLaMA-3 agents. In its present form, the detection capability is learned from labeled source files and is therefore bounded by the coverage and fidelity of those labels; it does not encode a vulnerability taxonomy, and it does not analyze compiled artifacts such as native extensions or bytecode.

Generalization to other ecosystems, such as npm or Maven, will require adapting the extractor to different layouts and build files and may require retraining on language-specific corpora. Ecosystem-specific installation and packaging conventions would need to be modeled explicitly, and the current implementation does not perform transitive dependency analysis or utilize package metadata signals, which limits the coverage of supply-chain attack surfaces. Experiments were executed locally on a MacBook Pro M3 using the MPS backend; therefore, the reported latency and throughput reflect this environment and may differ on other hardware. Differences in backend scheduling and tokenizer throughput can influence wall-clock measurements, and the system does not implement distributed or batched inference. Prompts for non-classifier agents are manually crafted, which introduces design bias and may not transfer unchanged to substantially different package structures. Systematic prompt sensitivity analysis and automated prompt tuning are not included. The aggregation policy favors recall at the package level; large packages containing rare but benign patterns that resemble learned malicious signals may yield false positives. Alternative thresholding or policy variants for different operating points are not explored here. 

\revised{
	The system neither performs dynamic execution nor emulates network activity. Moreover, it introduces no side effects associated with sandbox installation. Efficiency measurements in Table~\ref{tab:efficiency} reflect the behaviour of the current prototype and outline its operational cost profile under the evaluated configuration.
}

\subsubsection{Threats to Validity.}

\begin{itemize}
	\item \textit{Internal validity} can be affected by the design of prompts, the choices of hyperparameters for fine-tuning, and the specific model checkpoints used. We mitigate sources of randomness by performing package-level splits to avoid leakage across partitions, repeating all experiments with fixed seeds, pinning library versions, and using identical train–test splits across methods. We also apply McNemar’s test for paired comparisons and compute confidence intervals by bootstrap. Prompts for each agent are kept fixed across runs and are released in the replication package together with the exact seeds and environment specification, which supports reproducibility and auditability.
	
	\item \textit{External validity} is limited by the composition of D$_1$ and D$_2$ and by the time of data collection. The prevalence and style of malicious packages can shift over time, and class imbalance may differ across application settings. Although benign samples are filtered against public advisories and inspected, residual label noise may remain. The study focuses on PyPI; however, results may not be directly applicable without adaptation to ecosystems with different packaging conventions and installer practices. These factors bound the generality of the conclusions.
	
	\item \textit{Construct validity} reflects the use of learned code-level indicators as a proxy for malicious intent. The classifier operates on source files and does not observe runtime behavior, installation effects, or environment-dependent triggers, which can lead to mismatches between predicted intent and actual execution. Metrics such as accuracy, precision, recall, F1, and balanced accuracy capture correctness on the defined tasks, yet they do not fully characterize operational risk or downstream impact. The reported findings should therefore be interpreted within the defined datasets and protocol.
\end{itemize}
	
\section{Related Work}
\label{sec:related}

This section reviews related work in four complementary areas. We first discuss research on detecting malicious Python packages, followed by machine learning approaches for malware and vulnerability detection. We then examine recent advances in applying LLM-based multi-agent systems to software engineering tasks, and finally situate malicious package detection within the broader literature on software supply chain security in package ecosystems. Together, these strands provide the context for our study and clarify how our contribution aligns with and extends existing research.


\subsection{Detection of malicious PyPI Packages.} 

The increasing reliance on open-source repositories such as PyPI has introduced significant attack surfaces for malicious actors. The injection of harmful payloads into standard configuration files, most notably \texttt{setup.py}, has emerged as a recurring threat vector. Adversaries frequently rely on typosquatting, obfuscation, or misuse of dependency specifications to trick developers into installing malicious packages. These forms of attack are often subtle and evasive, posing substantial challenges for conventional rule-based or static analysis tools \citep{guo2023empirical,liang2021malicious,ruohonen2021large, nguyen2021adversarial}.

Prior studies have sought to address this challenge through different strategies. Static analysis approaches, such as large-scale scans of PyPI \citep{ruohonen2021large}, can reveal suspicious imports, command executions, or encoded payloads, but often suffer from high false positive rates and limited robustness against obfuscation. Tools such as Bandit4Mal and VirusTotal have been empirically shown to misclassify large proportions of samples, with false positives exceeding 80\% and false negatives surpassing 50\% in some benchmarks \citep{vu2022benchmark,nettersheim2024utilizing}. Dynamic analysis methods attempt to capture malicious behaviors by executing packages in sandboxed environments, as illustrated in prior work on adversarial injections in Python ecosystems \citep{liang2021malicious}, yet these approaches are resource-intensive and remain vulnerable to evasion through environment checks, delayed payload triggers, or conditional logic.

Learning-based strategies have recently been explored as more adaptive alternatives. Liang \etal proposed MPHunter, a clustering-based method that groups similar packages to identify anomalous outliers \citep{liang2021malicious}. This approach scales better than rule-based heuristics but remains sensitive to surface-level similarities; adversaries can still disguise malicious logic using obfuscation or mimicry of benign structures. More recently, Zhang \etal demonstrated a behavior-sequence model capable of detecting malicious packages across ecosystems such as npm and PyPI using a single model trained on behavioral signals \citep{zhang2023malicious}. While this expands generalization, such methods depend heavily on the availability of representative execution traces. At the ecosystem level, Alfadel \etal \citep{alfadel2023empirical} and Nguyen \etal \citep{nguyen2021adversarial} have highlighted that malicious packages not only threaten end users but also exploit weaknesses in dependency resolution and API recommendation systems, reinforcing the need for holistic supply chain defenses.

Although these efforts provide important baselines, they share limitations in scalability, robustness to obfuscation, and interpretability. Signature-based and static methods are brittle to novel attacks, clustering-based approaches can overlook semantic signals, and dynamic techniques are expensive to operate at the ecosystem scale. In contrast, our work investigates the feasibility of using large language models not as monolithic classifiers but as role-specialized agents that collaboratively curate inputs, perform classification, and aggregate package-level verdicts. By distributing responsibilities across independent yet coordinated agents, \tool leverages semantic reasoning while maintaining reproducibility and interpretability, offering a complementary path beyond traditional detection pipelines.

\subsection{Malware and Vulnerability Detection with Machine Learning}

Before the advent of LLMs, a large body of research investigated the use of classical and deep learning techniques for detecting malware and security vulnerabilities in source code. These approaches typically represent code through syntactic or semantic abstractions such as abstract syntax trees (ASTs), control-flow graphs (CFGs), or data-flow graphs (DFGs), which are then encoded into feature vectors for machine learning models. Early work applied traditional classifiers such as support vector machines (SVMs) and random forests to handcrafted code features, demonstrating moderate success in vulnerability prediction but limited scalability to unseen projects \citep{Chakraborty2020262,zhou2019devign}.

With the rise of deep learning, more expressive models were introduced. Graph-based neural networks have been applied to ASTs and program dependence graphs for vulnerability detection, capturing structural and semantic relations beyond surface syntax \citep{zhou2019devign}. Similarly, sequence-based architectures using RNNs, LSTMs, or Transformers have been trained directly on code tokens or learned embeddings, showing improved generalization across vulnerability datasets \citep{li2018vuldeepecker,allamanis2018survey}. Several benchmark datasets, including Devign \citep{zhou2019devign} and Big-Vul \citep{fan2020bigvul}, have enabled systematic evaluation of these approaches. Despite their effectiveness, these models are often domain-specific, require extensive feature engineering or graph construction, and struggle with adversarial obfuscation.

Parallel research has applied machine learning to malware detection at the binary and package level. Traditional malware classifiers leverage opcode sequences, API call graphs, or byte-level n-grams combined with supervised classifiers \citep{raff2018eating,liang2021malicious}. More recent work integrates deep neural networks for end-to-end malware detection, though these methods remain vulnerable to evasion through obfuscation, packing, or adversarial perturbations. Studies such as EC2 \citep{Chakraborty2020262} highlight that ensemble techniques can improve detection rates across malware families but still face challenges in interpretability and cross-platform generalization.

While these approaches laid the foundation for learning-based vulnerability and malware detection, they are limited in extensibility and transparency. Most systems train a monolithic model to perform end-to-end classification, which makes error analysis difficult and often reduces robustness to obfuscation or distributional shifts. In contrast, our work leverages LLMs within a modular multi-agent framework, where classification is complemented by input curation and package-level aggregation. This design preserves the advantages of supervised learning while adding interpretability, reproducibility, and scalability in the context of malicious package detection.


\subsection{Software Supply Chain Security in Package Ecosystems}

The problem of malicious packages is part of the broader challenge of securing the software supply chain. Recent incidents such as the SolarWinds compromise and large-scale injections of malicious dependencies in ecosystems like npm and PyPI have underscored the systemic risks that arise when attackers exploit trust relationships in package distribution networks \citep{ohm2020backstabber}. Research in this space has addressed multiple layers of the supply chain, including dependency management, vulnerability propagation, and ecosystem governance.

Empirical studies have shown that dependency networks in ecosystems such as PyPI, npm, and Maven expose substantial transitive risk: a single compromised package can affect thousands of downstream projects. Taxonomies of supply chain attacks \citep{ohm2020backstabber} highlight that injection through public repositories is only one of several attack vectors, alongside compromised build pipelines, stolen credentials, and malicious updates. Detection strategies, therefore, include dependency risk scoring, provenance verification, and anomaly detection across package metadata and version histories \citep{zahan2023software}.

Despite this progress, existing approaches face challenges in scalability and robustness against adversarial behavior. Signature-based and metadata-driven defenses can be bypassed by subtle obfuscation or typosquatting, while provenance frameworks remain limited by adoption barriers and ecosystem heterogeneity. In this context, our work on \tool complements existing lines of defense by focusing specifically on the semantic detection of malicious logic embedded in packages, and by investigating how modular LLM-based multi-agent systems can contribute to supply chain security with interpretable and auditable decision making.


\subsection{LLM-based MAS in Software Engineering} 

Recent work has begun to explore the application of multi-agent systems powered by large language models in software engineering tasks. These systems decompose complex activities into coordinated agents that collaborate via structured prompts, often outperforming monolithic pipelines. For example, CoRe \citep{10.1145/3691620.3695291} is a collaborative MAS for code reviewer recommendation, where agents leverage semantic representations of pull requests and reviewer expertise to improve reviewer assignment. Metagente \citep{nguyen2025teamworkmakesdreamwork} employs a multi-agent prompt optimization framework for summarizing GitHub README files, demonstrating improvements in text quality via ROUGE-based feedback. Other frameworks such as CAMEL \citep{li2023camel}, AgentVerse \citep{chen2023agentverse}, and CrewAI \citep{taulli2025crewai} illustrate the broader potential of role-specialized agents for cooperative reasoning, planning, and task execution.

Applications of MAS in software engineering extend beyond recommendation and summarization. DocAgent \citep{yang2025docagent} applies a multi-agent architecture to automated code documentation generation, while Arcadinho \etal \citep{arcadinho2024automated} demonstrate the use of agent collaboration for test generation and evaluation of LLM-powered tools. Ronanki \etal \citep{ronanki2023investigating} investigate conversational agents for requirements elicitation, showing that task decomposition improves coverage and reduces ambiguity. Beyond task-specific contributions, several surveys have recently mapped this emerging space. He \etal \citep{10.1145/3712003} review LLM-based MAS across the software lifecycle and articulate open challenges. Chen \etal \citep{chen2024survey} and Liu \etal \citep{liu2024llmagentsurvey} provide complementary surveys of multi-agent architectures and methodologies, highlighting challenges in coordination, evaluation, and communication protocols. At the systems level, Lee \etal \citep{lee2025reliable} examine strategies for reliable decision-making in MAS, comparing aggregation schemes such as majority voting, feedback loops, and arbitration. Becattini \citep{becattini2025sallma} proposes SALLMA, a reference architecture for LLM-based multi-agent systems, emphasizing modular orchestration and contextual memory management.

Altogether, this body of research demonstrates that LLM-driven MAS are versatile and increasingly applied across diverse SE tasks, from recommendation to documentation to testing. However, the majority of these contributions focus on productivity and collaboration rather than adversarial or security-critical contexts. In contrast, our work positions multi-agent orchestration for malicious package detection, where robustness, reproducibility, and interpretability are central. By combining input curation, file-level classification, and conservative package-level aggregation, \tool provides an instance of MAS applied to software supply chain security. This area has received limited attention in the current literature.

\section{Conclusion and Future Work}
\label{sec:conclusion}

This paper introduced \tool, a multi-agent system that leverages large language models to detect malicious Python packages through coordinated, role-specialized reasoning. The architecture departs from monolithic classification pipelines by decomposing the analysis into autonomous agents, each responsible for a semantically distinct task and orchestrated through natural language interaction. In our configuration, lightweight agents curate inputs and aggregate evidence, while a fine-tuned classifier operates at the file level. This separation improves stability under fixed seeds, keeps inputs within context limits, and yields auditable package-level decisions that can be inspected and replayed.

Experimental results show that \tool consistently outperforms competitive baselines, including \MP, RAG-based methods, and single-agent LLMs, with clear advantages in detecting obfuscated or sparsely expressed threats. 
Overall, the study demonstrates the feasibility of distributed LLM reasoning for malicious code detection and underscores the broader potential of multi-agent systems in secure software engineering. Beyond surpassing competitive baselines, the modular architecture enhances transparency, interpretability, and extensibility, as intermediate outputs and rationales can be audited and additional agents can be integrated without retraining the entire pipeline.

Future work will explore scaling the architecture and formalizing the design of roles and hand-offs, including ablation studies to quantify the individual contributions of input curation, file-level classification, and verdict aggregation. We plan to extend the empirical scope to larger and time-evolving corpora, as well as to other ecosystems such as npm or Maven, with appropriate adaptations of the extractor and training data. Another direction is the integration of dynamic and structural program features into the reasoning process, together with a more comprehensive efficiency model that reports latency, throughput, memory usage, and token consumption. Finally, we will investigate explanation quality by linking classifier rationales to concrete code regions and by evaluating explanation fidelity and stability across seeds, to improve interpretability and support audit and error analysis in security-critical contexts.

%
%
%
%
%
%
%


	\section*{Acknowledgments}
	This paper has been partially supported by the MOSAICO project (Management, Orchestration and Supervision of AI-agent COmmunities for reliable AI in software engineering) that has received funding from the European Union under the Horizon Research and Innovation Action (Grant Agreement No. 101189664). The work has been partially supported by the EMELIOT national research project, which has been funded by the MUR under the PRIN 2020 program (Contract 2020W3A5FY). It has been also partially supported by the European Union--NextGenerationEU through the Italian Ministry of University and Research, Projects PRIN 2022 PNRR \emph{``FRINGE: context-aware FaiRness engineerING in complex software systEms''} grant n. P2022553SL. We acknowledge the Italian ``PRIN 2022'' project TRex-SE: \emph{``Trustworthy Recommenders for Software Engineers,''} grant n. 2022LKJWHC. \revised{We thank the anonymous reviewers for their valuable comments and suggestions that helped us improve the manuscript.}



%
	\balance
	\bibliographystyle{abbrvnat}
	\bibliography{main}

@book{DBLP:books/sp/rsse2014,
	editor       = {Martin P. Robillard and
	Walid Maalej and
	Robert J. Walker and
	Thomas Zimmermann},
	title        = {Recommendation Systems in Software Engineering},
	publisher    = {Springer},
	year         = {2014},
	url          = {https://doi.org/10.1007/978-3-642-45135-5},
	doi          = {10.1007/978-3-642-45135-5},
	isbn         = {978-3-642-45134-8},
	bibsource    = {dblp computer science bibliography, https://dblp.org}
}

@misc{YARAForge2024,
	title        = {YARA Forge - A Rule-Sharing Platform for YARA},
	author       = {{YARA HQ}},
	year         = 2024,
	url          = {https://github.com/YARAHQ/yara-forge},
	note         = {Accessed: 2025-02-20}
}

@ARTICLE{Chakraborty2020262,
	author = {Chakraborty, Tanmoy and Pierazzi, Fabio and Subrahmanian, V.S.},
	title = {EC2: Ensemble Clustering and Classification for Predicting Android Malware Families},
	year = {2020},
	journal = {IEEE Transactions on Dependable and Secure Computing},
	volume = {17},
	number = {2},
	pages = {262 – 277},
	doi = {10.1109/TDSC.2017.2739145},
	url = {https://www.scopus.com/inward/record.uri?eid=2-s2.0-85028505970&doi=10.1109%2fTDSC.2017.2739145&partnerID=40&md5=43932ce129696a625bbb17be7f844d38},
	type = {Article},
	publication_stage = {Final},
	source = {Scopus},
	note = {Cited by: 93; All Open Access, Green Open Access}
}

@inproceedings{10.1145/3691620.3695492,
	title        = {SpiderScan: Practical Detection of Malicious NPM Packages Based on Graph-Based Behavior Modeling and Matching},
	author       = {Huang, Yiheng and Wang, Ruisi and Zheng, Wen and Zhou, Zhuotong and Wu, Susheng and Ke, Shulin and Chen, Bihuan and Gao, Shan and Peng, Xin},
	year         = 2024,
	booktitle    = {Proceedings of the 39th IEEE/ACM International Conference on Automated Software Engineering},
	location     = {Sacramento, CA, USA},
	publisher    = {Association for Computing Machinery},
	address      = {New York, NY, USA},
	series       = {ASE '24},
	pages        = {1146–1158},
	doi          = {10.1145/3691620.3695492},
	isbn         = 9798400712487,
	url          = {https://doi.org/10.1145/3691620.3695492},
	numpages     = 13
}

@inproceedings{10.1145/3510003.3510146,
	title        = {Natural attack for pre-trained models of code},
	author       = {Yang, Zhou and Shi, Jieke and He, Junda and Lo, David},
	year         = 2022,
	booktitle    = {Proceedings of the 44th International Conference on Software Engineering},
	location     = {Pittsburgh, Pennsylvania},
	publisher    = {Association for Computing Machinery},
	address      = {New York, NY, USA},
	series       = {ICSE '22},
	pages        = {1482–1493},
	doi          = {10.1145/3510003.3510146},
	isbn         = 9781450392211,
	url          = {https://doi.org/10.1145/3510003.3510146},
	numpages     = 12
}

@inproceedings{9647791,
	title        = {{ A Large-Scale Security-Oriented Static Analysis of Python Packages in PyPI }},
	author       = {Ruohonen, Jukka and Hjerppe, Kalle and Rindell, Kalle},
	year         = 2021,
	month        = {Dec},
	booktitle    = {2021 18th International Conference on Privacy, Security and Trust (PST)},
	publisher    = {IEEE Computer Society},
	address      = {Los Alamitos, CA, USA},
	volume       = {},
	pages        = {1--10},
	doi          = {10.1109/PST52912.2021.9647791},
	issn         = {},
	url          = {https://doi.ieeecomputersociety.org/10.1109/PST52912.2021.9647791}
}

@article{10.1145/3705304,
	title        = {Killing Two Birds with One Stone: Malicious Package Detection in NPM and PyPI using a Single Model of Malicious Behavior Sequence},
	author       = {Zhang, Junan and Huang, Kaifeng and Huang, Yiheng and Chen, Bihuan and Wang, Ruisi and Wang, Chong and Peng, Xin},
	year         = 2025,
	month        = apr,
	journal      = {ACM Trans. Softw. Eng. Methodol.},
	publisher    = {Association for Computing Machinery},
	address      = {New York, NY, USA},
	volume       = 34,
	number       = 4,
	doi          = {10.1145/3705304},
	issn         = {1049-331X},
	url          = {https://doi.org/10.1145/3705304},
	issue_date   = {May 2025},
	articleno    = 104,
	numpages     = 28
}

@inproceedings{10.1145/3691620.3695291,
	title        = {Unity Is Strength: Collaborative LLM-Based Agents for Code Reviewer Recommendation},
	author       = {Wang, Luqiao and Zhou, Yangtao and Zhuang, Huiying and Li, Qingshan and Cui, Di and Zhao, Yutong and Wang, Lu},
	year         = 2024,
	booktitle    = {Proceedings of the 39th IEEE/ACM ASE},
	publisher    = {ACM},
	address      = {New York, NY, USA},
	series       = {ASE '24},
	pages        = {2235–2239},
	doi          = {10.1145/3691620.3695291},
	isbn         = 9798400712487,
	url          = {https://doi.org/10.1145/3691620.3695291},
	numpages     = 5
}

@inproceedings{10298315,
	title        = {A Needle is an Outlier in a Haystack: Hunting Malicious PyPI Packages with Code Clustering},
	author       = {Liang, Wentao and Ling, Xiang and Wu, Jingzheng and Luo, Tianyue and Wu, Yanjun},
	year         = 2023,
	booktitle    = {2023 38th IEEE/ACM International Conference on Automated Software Engineering (ASE)},
	volume       = {},
	number       = {},
	pages        = {307--318},
	doi          = {10.1109/ASE56229.2023.00085}
}

@inproceedings{ibiyo2025detectingmalicioussourcecode,
	title        = {{Detecting Malicious Source Code in PyPI Packages with LLMs: Does RAG Come in Handy?}},
	author       = {Ibiyo, Motunrayo and Louangdy, Thinakone and Nguyen, Phuong T. and Di Sipio, Claudio and Di Ruscio, Davide},
	year         = 2025,
	booktitle    = {Proceedings of the 29th International Conference on Evaluation and Assessment in Software Engineering},
	location     = {Istanbul, Tur},
	publisher    = {Association for Computing Machinery},
	address      = {New York, NY, USA},
	series       = {EASE '25},
	doi          = {10.1145/3593434.3593448},
	isbn         = 9798400700446,
	url          = {https://doi.org/10.1145/3593434.3593448}
}

@article{chen2021evaluating,
	title        = {Evaluating large language models trained on code},
	author       = {Chen, Mark and Tworek, Jerry and Jun, Heewoo and Yuan, Qiming and Pinto, Henrique Ponde De Oliveira and Kaplan, Jared and Edwards, Harri and Burda, Yuri and Joseph, Nicholas and Brockman, Greg and others},
	year         = 2021,
	journal      = {arXiv preprint arXiv:2107.03374}
}

@article{ahmad2021unified,
	title        = {Unified pre-training for program understanding and generation},
	author       = {Ahmad, Wasi Uddin and Chakraborty, Saikat and Ray, Baishakhi and Chang, Kai-Wei},
	year         = 2021,
	journal      = {arXiv preprint arXiv:2103.06333}
}

@article{wang2023can,
	title        = {Can ChatGPT defend its belief in truth? evaluating LLM reasoning via debate},
	author       = {Wang, Boshi and Yue, Xiang and Sun, Huan},
	year         = 2023,
	journal      = {arXiv preprint arXiv:2305.13160}
}

@inproceedings{ohm2020backstabber,
	title        = {Backstabber’s knife collection: A review of open source software supply chain attacks},
	author       = {Ohm, Marc and Plate, Henrik and Sykosch, Arnold and Meier, Michael},
	year         = 2020,
	booktitle    = {Detection of Intrusions and Malware, and Vulnerability Assessment: 17th International Conference, DIMVA 2020, Lisbon, Portugal, June 24--26, 2020, Proceedings 17},
	pages        = {23--43},
	organization = {Springer}
}

@article{alfadel2023empirical,
	title        = {Empirical analysis of security vulnerabilities in python packages},
	author       = {Alfadel, Mahmoud and Costa, Diego Elias and Shihab, Emad},
	year         = 2023,
	journal      = {Empirical Software Engineering},
	publisher    = {Springer},
	volume       = 28,
	number       = 3,
	pages        = 59
}

@article{chen2023agentverse,
	title        = {Agentverse: Facilitating multi-agent collaboration and exploring emergent behaviors in agents},
	author       = {Chen, Weize and Su, Yusheng and Zuo, Jingwei and Yang, Cheng and Yuan, Chenfei and Qian, Chen and Chan, Chi-Min and Qin, Yujia and Lu, Yaxi and Xie, Ruobing and others},
	year         = 2023,
	journal      = {arXiv preprint arXiv:2308.10848},
	volume       = 2,
	number       = 4,
	pages        = 6
}

@incollection{taulli2025crewai,
	title        = {CrewAI},
	author       = {Taulli, Tom and Deshmukh, Gaurav},
	year         = 2025,
	booktitle    = {Building Generative AI Agents},
	publisher    = {Springer},
	pages        = {103--145}
}

@inproceedings{guo2023empirical,
	title        = {An empirical study of malicious code in pypi ecosystem},
	author       = {Guo, Wenbo and Xu, Zhengzi and Liu, Chengwei and Huang, Cheng and Fang, Yong and Liu, Yang},
	year         = 2023,
	booktitle    = {2023 38th IEEE/ACM International Conference on Automated Software Engineering (ASE)},
	pages        = {166--177},
	organization = {IEEE}
}

@inproceedings{liang2021malicious,
	title        = {Malicious packages lurking in user-friendly python package index},
	author       = {Liang, Genpei and Zhou, Xiangyu and Wang, Qingyu and Du, Yutong and Huang, Cheng},
	year         = 2021,
	booktitle    = {2021 IEEE 20th International Conference on Trust, Security and Privacy in Computing and Communications (TrustCom)},
	pages        = {606--613},
	organization = {IEEE}
}

@inproceedings{nguyen2021adversarial,
	title        = {Adversarial attacks to API recommender systems: time to wake up and smell the coffee?},
	author       = {Nguyen, Phuong T. and {Di Sipio}, Claudio and {Di Rocco}, Juri and {Di Penta}, Massimiliano and {Di Ruscio}, Davide},
	year         = 2022,
	booktitle    = {Proceedings of the 36th IEEE/ACM International Conference on Automated Software Engineering},
	location     = {Melbourne, Australia},
	publisher    = {IEEE Press},
	series       = {ASE '21},
	pages        = {253–265},
	doi          = {10.1109/ASE51524.2021.9678946},
	isbn         = 9781665403375,
	url          = {https://doi.org/10.1109/ASE51524.2021.9678946},
	numpages     = 13
}

@article{10.1145/3712003,
	title        = {LLM-Based Multi-Agent Systems for Software Engineering: Literature Review, Vision and the Road Ahead},
	author       = {He, Junda and Treude, Christoph and Lo, David},
	year         = 2025,
	month        = jan,
	journal      = {ACM Trans. Softw. Eng. Methodol.},
	publisher    = {Association for Computing Machinery},
	address      = {New York, NY, USA},
	doi          = {10.1145/3712003},
	issn         = {1049-331X},
	url       = {https://doi.org/10.1145/3712003}
}

@inbook{nguyen2025teamworkmakesdreamwork,
	author = {Nguyen, Duc S. H. and Truong, Bach G. and Nguyen, Phuong T. and {Di Rocco}, Juri and {Di Ruscio}, Davide},
	title = {Teamwork makes the dream work: LLMs-Based Agents for GitHub README.MD Summarization},
	year = {2025},
	isbn = {9798400712760},
	publisher = {Association for Computing Machinery},
	address = {New York, NY, USA},
	url = {https://doi.org/10.1145/3696630.3728511},
	booktitle = {Proceedings of the 33rd ACM International Conference on the Foundations of Software Engineering},
	pages = {621–625},
	numpages = {5}
}

@article{pembury2020effective,
	title={Effective use of the McNemar test},
	author={Pembury Smith, Matilda QR and Ruxton, Graeme D},
	journal={Behavioral Ecology and Sociobiology},
	volume={74},
	pages={1--9},
	year={2020},
	publisher={Springer}
}

@misc{lampsjss2025,
  author       = {Muhammad Umar {Zeshan} and Motunrayo Ibiyo and Phuong T. Nguyen and Claudio {Di Sipio} and Davide {Di Ruscio}},
  title        = {{Replication Package: ``Many Hands Make Light Work: An LLM-based Multi Agent System for Detecting Malicious PyPI Packages''}},
  howpublished = {\url{https://github.com/muzeshan/lamps-jss}},
  note         = {Accessed: September 14, 2025},
  year         = {2025}
}

@misc{li2023camel,
      title={CAMEL: Communicative Agents for "Mind" Exploration of Large Language Model Society}, 
      author={Guohao Li and Hasan Abed Al Kader Hammoud and Hani Itani and Dmitrii Khizbullin and Bernard Ghanem},
      year={2023},
      eprint={2303.17760},
      archivePrefix={arXiv},
      primaryClass={cs.AI},
      url={https://arxiv.org/abs/2303.17760}, 
}

@article{yang2025docagent,
	title={DocAgent: A Multi-Agent System for Automated Code Documentation Generation},
	author={Yang, Dayu and Simoulin, Antoine and Qian, Xin and Liu, Xiaoyi and Cao, Yuwei and Teng, Zhaopu and Yang, Grey},
	journal={arXiv preprint arXiv:2504.08725},
	year={2025}
}

@article{arcadinho2024automated,
	title={Automated test generation to evaluate tool-augmented LLMs as conversational AI agents},
	author={Arcadinho, Samuel and Apar{\'\i}cio, David and Almeida, Mariana},
	journal={arXiv preprint arXiv:2409.15934},
	year={2024}
}

@inproceedings{ronanki2023investigating,
	title={Investigating ChatGPT’s potential to assist in requirements elicitation processes},
	author={Ronanki, Krishna and Berger, Christian and Horkoff, Jennifer},
	booktitle={2023 49th Euromicro Conference on Software Engineering and Advanced Applications (SEAA)},
	pages={354--361},
	year={2023},
	organization={IEEE}
}

@inproceedings{ruohonen2021large,
	author    = {Jukka Ruohonen and Kalle Hjerppe and Kalle Rindell},
	title     = {A Large-Scale Security-Oriented Static Analysis of Python Packages in PyPI},
	booktitle = {2021 18th International Conference on Privacy, Security and Trust (PST)},
	year      = {2021},
	pages     = {1--10},
	publisher = {IEEE},
	doi       = {10.1109/PST52912.2021.9647791}
}

@inproceedings{nettersheim2024utilizing,
	author    = {Florian Nettersheim and Stephan Arlt and Michael Rademacher},
	title     = {Utilizing {DNS} and VirusTotal for Automated Ad-Malware Detection},
	booktitle = {Web Engineering: 24th International Conference, ICWE 2024, Tampere, Finland, June 17--20, 2024, Proceedings},
	series    = {Lecture Notes in Computer Science},
	volume    = {14629},
	pages     = {389--392},
	publisher = {Springer},
	year      = {2024},
	doi       = {10.1007/978-3-031-62362-2_31}
}

@misc{vu2022benchmark,
	author       = {Duc{-}Ly Vu and Zachary Newman and John Speed Meyers},
	title        = {A Benchmark Comparison of Python Malware Detection Approaches},
	year         = {2022},
	eprint       = {2209.13288},
	archivePrefix= {arXiv},
	primaryClass = {cs.CR},
	url          = {https://arxiv.org/abs/2209.13288}
}

@article{liu2024llmagentsurvey,
	author    = {Junwei Liu and Kaixin Wang and Yixuan Chen and Xin Peng and Zhenpeng Chen and Lingming Zhang and Yiling Lou},
	title     = {Large Language Model-Based Agents for Software Engineering: A Survey},
	journal   = {arXiv preprint},
	year      = {2024},
	url       = {https://arxiv.org/abs/2409.02977}
}

@inproceedings{lee2025reliable,
	author    = {Xian Yeow Lee and Shunichi Akatsuka and Lasitha Vidyaratne and Aman Kumar and Ahmed Farahat and Chetan Gupta},
	title     = {Reliable Decision-Making for Multi-Agent LLM Systems},
	booktitle = {Proceedings of the 2025 Workshop on Multi-Agent Systems},
	year      = {2025},
	url       = {https://multiagents.org/2025_artifacts/reliable_decision_making_for_multi_agent_llm_systems.pdf}
}

@inproceedings{zhou2019devign,
	author    = {Yaqin Zhou and Shangqing Liu and Jingkai Siow and Xiaoning Du and Yang Liu},
	title     = {Devign: Effective Vulnerability Identification by Learning Comprehensive Program Semantics via Graph Neural Networks},
	booktitle = {Advances in Neural Information Processing Systems 32 (NeurIPS 2019)},
	year      = {2019},
	url       = {https://arxiv.org/pdf/1909.03496.pdf}
}

@inproceedings{li2018vuldeepecker,
	author    = {Zhen Li and Deqing Zou and Shouhuai Xu and Xinyu Ou and Hai Jin and Sujuan Wang and Zhijun Deng and Yuyi Zhong},
	title     = {VulDeePecker: A Deep Learning-Based System for Vulnerability Detection},
	booktitle = {Proceedings of the Network and Distributed System Security Symposium (NDSS)},
	year      = {2018},
	doi       = {10.14722/NDSS.2018.23158},
	url       = {https://www.ndss-symposium.org/wp-content/uploads/2018/02/ndss2018_03A-2_Li_paper.pdf}
}

@article{allamanis2018survey,
	author    = {Miltiadis Allamanis and Earl T. Barr and Premkumar Devanbu and Charles Sutton},
	title     = {A Survey of Machine Learning for Big Code and Naturalness},
	journal   = {ACM Computing Surveys},
	volume    = {51},
	number    = {4},
	articleno = {81},
	year      = {2018},
	doi       = {10.1145/3212695}
}

@inproceedings{fan2020bigvul,
	author    = {Jiahao Fan and Yi Li and Shaohua Wang and Tien N. Nguyen},
	title     = {A C/C++ Code Vulnerability Dataset with Code Changes and CVE Summaries},
	booktitle = {Proceedings of the 17th International Conference on Mining Software Repositories (MSR)},
	pages     = {508--512},
	year      = {2020},
	doi       = {10.1145/3379597.3387501}
}

@inproceedings{raff2018eating,
	author    = {Edward Raff and Jon Barker and Jared Sylvester and Robert Brandon and Bryan Catanzaro and Charles Nicholas},
	title     = {Malware Detection by Eating a Whole EXE},
	booktitle = {AAAI Workshop on Artificial Intelligence for Cyber Security (AICS)},
	year      = {2018},
	url       = {https://arxiv.org/abs/1710.09435}
}

@inproceedings{samaana2025machine,
  title={A Machine Learning-Based Approach For Detecting Malicious PyPI Packages},
  author={Samaana, Haya and Costa, Diego Elias and Shihab, Emad and Abdellatif, Ahmad},
  booktitle={Proceedings of the 40th ACM/SIGAPP Symposium on Applied Computing},
  pages={1617--1626},
  year={2025}
}

@inproceedings{choi2023codeprompt,
  title={CodePrompt: Task-agnostic prefix tuning for program and language generation},
  author={Choi, YunSeok and Lee, Jee-Hyong},
  booktitle={Findings of the Association for Computational Linguistics: ACL 2023},
  pages={5282--5297},
  year={2023}
}

@article{chen2024survey,
  title={A survey on llm-based multi-agent system: Recent advances and new frontiers in application},
  author={Chen, Shuaihang and Liu, Yuanxing and Han, Wei and Zhang, Weinan and Liu, Ting},
  journal={arXiv preprint arXiv:2412.17481},
  year={2024}
}

@article{chen2023autoagents,
  title={Autoagents: A framework for automatic agent generation},
  author={Chen, Guangyao and Dong, Siwei and Shu, Yu and Zhang, Ge and Sesay, Jaward and Karlsson, B{\"o}rje F and Fu, Jie and Shi, Yemin},
  journal={arXiv preprint arXiv:2309.17288},
  year={2023}
}

@inproceedings{zahan2023software,
  title={Software supply chain risk assessment framework},
  author={Zahan, Nusrat},
  booktitle={2023 IEEE/ACM 45th International Conference on Software Engineering: Companion Proceedings (ICSE-Companion)},
  pages={251--255},
  year={2023},
  organization={IEEE}
}

@inproceedings{becattini2025sallma,
  title={SALLMA: A Software Architecture for LLM-Based Multi-Agent Systems},
  author={Becattini, Marco and Verdecchia, Roberto and Vicario, Enrico},
  booktitle={2025 IEEE/ACM International Workshop New Trends in Software Architecture (SATrends)},
  pages={5--8},
  year={2025},
  organization={IEEE}
}

@article{zhang2023malicious,
  title={Malicious package detection in npm and pypi using a single model of malicious behavior sequence},
  author={Zhang, Junan and Huang, Kaifeng and Chen, Bihuan and Wang, Chong and Tian, Zhenhao and Peng, Xin},
  journal={arXiv preprint arXiv:2309.02637},
  year={2023}
}

\end{document}